\documentclass[twocolumn,fleqn,10pt]{wlpeerj} 
\usepackage{subcaption,booktabs}

\title{Robust inference of causality in high-dimensional dynamical processes from the Information Imbalance of distance ranks}
\author[1]{Vittorio Del Tatto}
\author[1]{Gianfranco Fortunato}
\author[1]{Domenica Bueti}
\author[1,2]{Alessandro Laio}
\affil[1]{Scuola Internazionale Superiore di Studi Avanzati, SISSA, via Bonomea 265, 34136 Trieste, Italy}
\affil[2]{ICTP, Strada Costiera 11, 34151 Trieste, Italy}

\begin{abstract}
We introduce an approach which allows detecting causal relationships between variables for which the time evolution is available.  
Causality is assessed by a variational scheme based on the Information Imbalance of distance ranks, a statistical test capable of inferring the relative information content of different distance measures.
We test whether the predictability of a putative driven system Y can be improved by incorporating information from a potential driver system X, without explicitly modeling the underlying dynamics and without the need to compute probability densities of the dynamic variables.
This framework makes causality detection possible even between high-dimensional systems where only few of the variables are known or measured. Benchmark tests on coupled chaotic dynamical systems demonstrate that our approach outperforms other model-free causality detection methods, successfully handling both unidirectional and bidirectional couplings.
We also show that the method can be used to robustly detect causality in human electroencephalography data. 
\end{abstract}

\begin{document}

\flushbottom
\maketitle
\thispagestyle{empty}

\section*{Introduction}
Discovering causal relationships among observable quantities has been inspiring and guiding scientific research from its dawn, as causality is at the very heart of physical phenomena and natural laws.
The definition of causality is far from univocal, with diverse frameworks rooted in distinct perspectives.
Granger's paradigm is based on ``predictive causality'' \cite{granger1969investigating}, while Pearl’s structural causal model \cite{pearl2009causality} builds on counterfactuals.
Determining causality from data collected without directly intervening on the system under study - namely, without performing interventional experiments where the causal variable is manipulated - is a challenging problem which received increasing attention over the last decades 
\cite{spirtes2016causal,mooij2016distinguishing,gendron2023survey}.
The use of purely observational data is the only option when experiments are unfeasible or unethical, such as in the case of medical studies that would create a real risk for patient's health, or Earth science researches that could alter delicate ecological balances, just to give a few examples. This motivated the development of statistical tests aimed specifically at inferring causal relationships in ``real world'' time-ordered data. These approaches are routinely employed 
in diverse fields, from  economics \cite{granger1969investigating, hoover2017causality} to ecology \cite{sugihara2012detecting}, Earth system sciences \cite{kretschmer2016using} and neuroscience \cite{friston2003dynamic, kaminski2001evaluating, bressler2011wiener}.
The common idea to all these methods is to compute statistical measures which are asymmetric under the exchange of the dynamic variables, in order to reflect the asymmetry of a putative causal coupling.
In this field some important conceptual and practical problems remain open and are still object of intense investigation.
In particular, the fact that real-world time series often emerge from complex underlying dynamics naturally brings to the necessity of methodologies dealing with high-dimensional data \cite{runge2018causal, runge2019inferring}.
Moreover, false positive detections represent a common yet crucial limitation even in low-dimensional scenarios \cite{krakovska2018comparison,yuan2022datadriven}.

From a historical perspective, the first quantitative criterion to measure causality dates back to the work of Wiener \cite{beckenbach1956modern}, who postulated that the prediction of a signal $Y$ can be improved by using the past information of a signal $X$ if $X$ is causal to $Y$.
Inspired by Wiener, Granger proposed to identify causal links in time series analysis with a vector autoregressive model assuming a linear dynamics \cite{granger1969investigating}.
Since then, several nonlinear generalizations of Granger's idea have been proposed \cite{Chen2004analyzing, marinazzo2006nonlinear, marinazzo2008kernel,barnett2015granger,krakovska2016testing,wang2022newmethod}.
In particular, the Extended Granger Causality test \cite{Chen2004analyzing} confines the linear approximation of the dynamics to local regions in the state space.
Causality can also be inferred by estimating a conditional mutual information named Transfer Entropy \cite{schreiber2000measuring,palus2001synchronization,palus2007directionality},
which is equivalent to Granger causality for Gaussian variables \cite{barnett2009granger}.
However, computing \emph{multivariate} probability distributions to evaluate mutual informations is challenging in high-dimensional systems, and in practice one is typically forced to work with conditional probabilities of a few variables at a time \cite{runge2012escaping}.
For this reason, alternative methods that do not require to compute the probability distributions of the dynamic variables are more appealing for real-world applications.
Among these, cross mapping methods rely on Takens' theorem \cite{takens1981dynamical}, which allows reconstructing a dynamical system's attractor - or rather a version capturing its main features, called \emph{shadow manifold} - using one-dimensional time series.
In particular, Convergent Cross Mapping \cite{sugihara2012detecting} evaluates the coupling strength $X\rightarrow Y$ by attempting a local reconstruction of the shadow manifold of $X$ from the shadow manifold of $Y$ and computing a correlation coefficient between the reconstructed and the ground-truth points.
Another cross mapping method, known as measure $L$ \cite{chicharro2009reliable}, employs a similar approach, but carries out the local reconstruction of the target manifold using ranks rather than distances.

In this work we introduce a causality detection method broadly based on Granger causality principles, namely that the cause occurs before the effect and the cause contains unique information about the effect.
We implement these principles using the Information Imbalance \cite{glielmo2022ranking}, a statistical measure designed to compare distance spaces and decide which space is more informative, without modelling the underlying dynamics.
By measuring distances between independent realizations of the same dynamical process, the method evaluates whether including the putative driver variables in the distance space built at time $t=0$ allows to better guess which pairs of trajectories will be the closest in the space of the driven variables at a future time $t=\tau$.
The information of the driver system is added to the space of the putative driven system using a variational approach, which allows probing the presence or absence of a coupling within a theoretically rigorous framework.
The distances in different spaces are compared by analyzing the statistics of distance ranks and examining how it is affected by the inclusion of potential causal variables in the distance definition.
Ranks can be easily obtained by ordering distances from the smallest to the largest, and the effort to compute them is not affected by the underlying dimensionality.

To assess the validity of the method we carried out tests on a variety of coupled dynamical systems with both unidirectional and bidirectional couplings, as well as on real-world time series from electroencephalography (EEG) experiments.
These tests led us to observe that our method 
allows recognizing with high statistical confidence  when  a causal coupling is \emph{absent}. Remarkably, we find that   the other approaches we tested fail systematically in this task, bringing to high probabilities of observing \emph{false positives}, namely of confusing the absence of causality with a condition of weak causal coupling. 
Our approach, besides strongly mitigating this problem, provides reliable results also when the dynamics system is \emph{high-dimensional}, such as the electrophysiological signal of a human brain, or a Lorenz 96 system \cite{lorenz2006predictability}.

\section*{The Information Imbalance Gain}
The Information Imbalance is a statistical measure introduced to compare the information content of two distances $d_A$ and $d_B$ defined on a set of points $\{x_i\}$ $(i=1,...,N)$, which allows assessing whether the distances are equivalent, independent or if one is more informative than the other
\cite{glielmo2022ranking}.


This measure is built on the idea that close points according to $d_A$ remain close in distance space $d_B$ when $d_A$ is informative with respect to $d_B$ or, equivalently, when the information carried by $d_B$ is also contained in $d_A$. 
We denote by $r_{ij}^A$ (resp. $r_{ij}^B$) the rank of  point $j$ with respect to point $i$ according to distance $d_A$ (resp. $d_B$), with the convention $r_{ii}^A = r_{jj}^B = 0$.
For example, $r_{ij}^A = 1$ if $j$ is the nearest neighbor of $i$ in space $A$.
The Information Imbalance from $A$ to $B$ is defined as the average rank according to distance $d_B$ restricted to points which are “close” according to $d_A$:
\begin{equation}\label{eq:info_imbalance}
    \Delta\left(d_A\rightarrow d_B\right) = \frac{2}{N}\langle r^B \mid r^A \le k \rangle\, = \frac{2}{N^2\, k} \sum_{\substack{i,j\\\text{s.t.}\, r_{ij}^A \le k}} r_{ij}^B \,.
\end{equation}
The parameter $k$ specifies the number of neighbors taken into account and generalizes the definition in Ref. \cite{glielmo2022ranking}, which assumes $k=1$.
This definition can also be interpreted as an asymptotic upper bound of a \emph{restricted mutual information}, that we define in the \emph{SI Appendix} (section 1).
The prefactor $2/N$ statistically confines this quantity between 0 and 1, which are the limits of $d_A$ being respectively maximally and minimally informative with respect to $d_B$.
As discussed in the \emph{SI Appendix} (section 2A), in this form the Information Imbalance is equivalent to the measure $L$ introduced by Chicharro and Andrzejak in Ref. \cite{chicharro2009reliable} for studying coupled dynamical systems. However, the way we apply this statistic to the problem of causal detection is substantially different, as discussed below.

\begin{figure*}
\centering
\includegraphics[width=14.5cm]{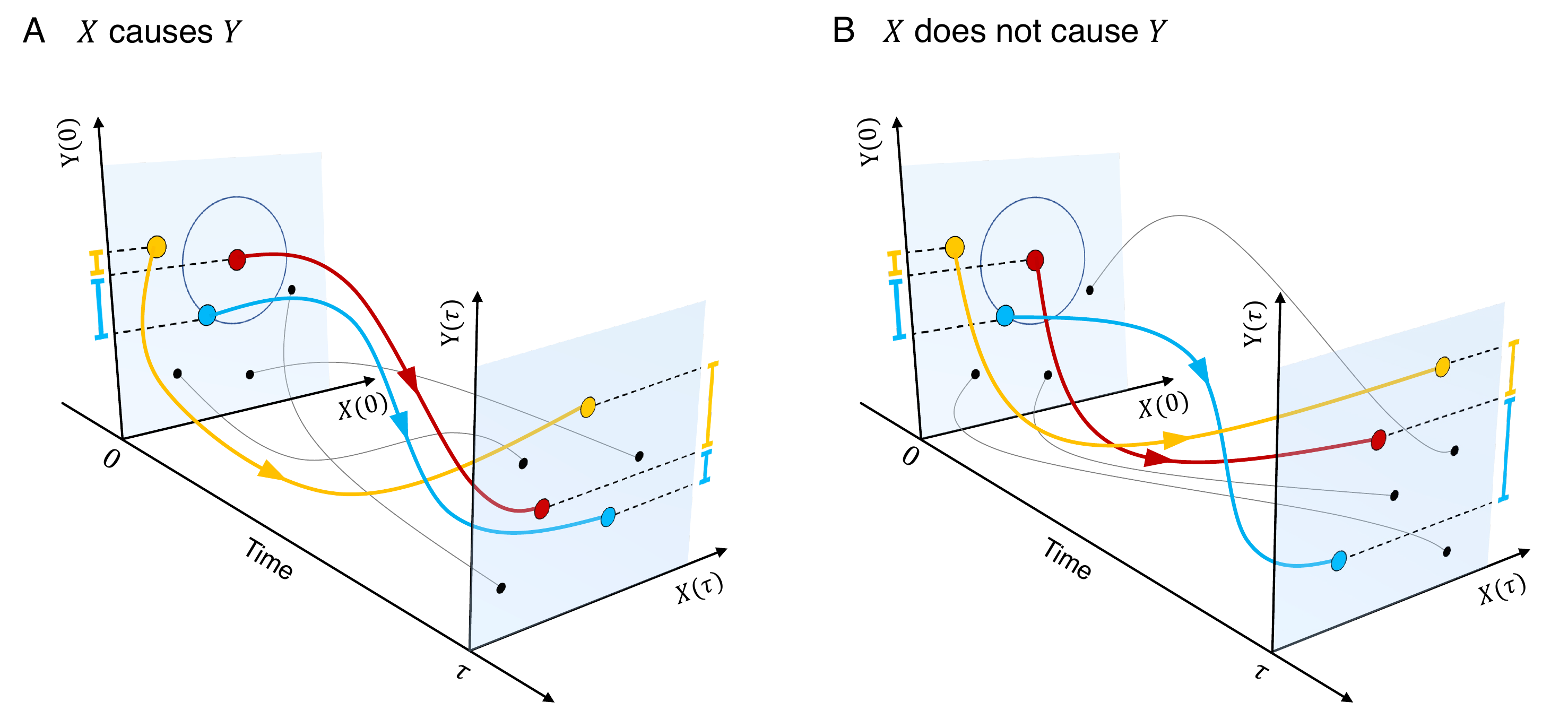}
\caption{Illustration of the method applied to the $X\rightarrow Y$ direction. Different lines represent independent realizations of a two-dimensional dynamical process. Both in (\emph{A}) and in (\emph{B}) the reference trajectory is depicted in red. At time 0 the blue trajectory is the closest in space $d_{X(0), Y(0)}$, while the yellow realization is the closest in the marginal space $d_{Y(0)}$. When we look at distances in the marginal space $d_{Y(\tau)}$, (\emph{A}) if $X$ causes $Y$ the closest curve is the blue one, which was the closest in the full dynamical space, while (\emph{B}) if $X$ does not cause $Y$ the yellow trajectory remains the closest, as the space $d_{Y(0)}$ already contains the maximal information to predict the state of $Y$ at time $\tau$.
Our method tries to predict the $k$ closest trajectories to each realization in the distance space of the putative driven system, assessing the prediction quality as the average of their true distance ranks.
The Imbalance Gain in Eq. (\ref{eq:rel_imbalance_gain}) allows comparing the predictions made in presence or in absence of the putative causal variables.}
\label{fig:fig_cartoon}
\end{figure*}

We here propose a variational approach, based on the Information Imbalance, to infer the presence of causal relationships among two sets of time-dependent variables, using no prior knowledge about the underlying dynamics.
To set the framework, let $X(t)$ and $Y(t)$ be vectors characterizing the states of two dynamical systems at time $t$, with components $x_\alpha(t)$ and $y_\beta(t)$ ($\alpha=1,...,D$; $\beta=1,...,D'$).
We suppose that all the components of $X$ (and, separately, of $Y$) are dynamically intertwined, namely that there are no proper subsets of coordinates of $X$ (resp, of $Y$) that are autonomous. This condition implies that the components of $X$ and $Y$ cannot be regrouped in three distinct systems $X'$, $Y'$ and $Z'$ such that $Z'$ is unidirectionally coupled both to $X'$ and $Y'$, namely a so-called common driver \cite{castro2023timeseries}.
The generalization of our approach in presence of a third \emph{observed} system $Z$, which may be a common driver of $X$ and $Y$, is presented and discussed in the \emph{SI Appendix} (section 3 and Fig. S1).
We do not address in this work the problem of \emph{unobserved} common drivers, in presence of which the methodology that we propose carries the risk of incorrectly identifying causal relationships.

Our approach is mainly benchmarked on dynamics which are smooth and deterministic, but it can be applied with no modification to stochastic processes (see section 4 and Fig. S2 in the \emph{SI Appendix}).
Since the Information Imbalance is computed over a set of points, we suppose to have access to multiple experiments $X^i(t)$ and $Y^i(t)$ ($i=1,...,N$), representing independent realizations of identical copies of the systems.
If the available data consist in a single stationary time series, an ensemble of multiple realizations can be constructed by dividing the trajectory in $N$ non-overlapping subtrajectories, which can be made efficaciously independent by increasing the time interval between the end of one trajectory and the beginning of the next.
In the following the distance spaces appearing in the Information Imbalance will be labeled by the coordinates employed to construct them. For example, we will use the notation $d_{X(0),Y(\tau)}$ to identify the following Euclidean distance between trajectories $i$ and $j$: $\left ( \|X^i(0)-X^j(0) \|^2+\|Y^i(\tau)-Y^j(\tau) \|^2 \right )^{\frac{1}{2}}.$

Our causality detection method relies on the intuition that if a dynamic variable $X$ causes another variable $Y$ and one attempts to make a prediction on the future of $Y$, a distance measure built using the present states of both $X$ and $Y$ will have more predictive power than a distance built using only $Y$. This idea is depicted in Fig. \ref{fig:fig_cartoon}.  
Formally, we postulate that if $X$ does not cause $Y$, then for any $\alpha>0$
\begin{equation}\label{eq:ineq}
    \Delta\left( \alpha \right) \doteq \,\Delta\big(d_{\alpha X(0),Y(0)}\rightarrow d_{Y\left(\tau\right)} \big) > \Delta\big( d_{Y(0)}\rightarrow d_{Y\left(\tau\right)} \big)\,.
\end{equation}
Indeed, if $Y$ is autonomous, adding information on the initial value of $X$ can only degrade the information on the future of $Y$.
If, instead, $X$  causes $Y$, adding information on $X$ will improve the predictability of the future of $Y$, and $\Delta\left( \alpha \right)$ will be minimized by some $\alpha>0$.
The parameter $\alpha$, as we show in the \emph{SI Appendix} (section 5 and Fig. S3), plays the role of a scaling parameter for the units of $X$ accounting  for the magnitude of the coupling strength.
$\tau$ is a positive parameter representing the time lag of the information transfer from the driving to the driven system.
Since this parameter only appears in the argument of the putative driven system $Y$, the approach is still valid when $Y$ is a dynamical system and $X$ is a static variable (for example a control variable chosen by an experimentalist), with the caveat that in this scenario causality can only be tested in direction $X\rightarrow Y$.
Using Eq. (\ref{eq:ineq}) we can assess the presence of causality by a variational scheme.
For this purpose we define the Imbalance Gain in direction $X\rightarrow Y$ as
\begin{equation}\label{eq:rel_imbalance_gain}
    \delta \Delta{(X\rightarrow Y) \doteq} \frac{\Delta(\alpha = 0) - \min_{\alpha}\Delta(\alpha)}{\Delta(\alpha = 0)}\,.
\end{equation}
Notice that $\delta \Delta {(X\rightarrow Y)}$ is by construction a positive definite quantity.
A value of $\delta \Delta {(X\rightarrow Y)} =0$ indicates that adding the information on the value of $X$ does not help predicting the future of $Y$, namely $Y$ is autonomous. If instead $\delta \Delta {(X\rightarrow Y)} > 0$, we infer that $X$ causes  $Y$.
In this second scenario, we will show that the value of the Imbalance Gain can be used to compare the strengths of different couplings.


Our approach can be viewed as a nonlinear and model-free generalization of Granger causality, as it examines the impact of introducing the supposed causal variable in the past on the predictability of the supposed caused variable in the future.

Ideally, the distance spaces appearing in Eq. (\ref{eq:ineq}) should be constructed using all the components of $X$ and $Y$. 
However, in real experiments it is common that not all the variables of each system are recorded. 
In this case, Takens' theorem \cite{takens1981dynamical} ensures that it is possible to recover the information of the missing coordinates by means of the time-delay embeddings of the known variables.
For example, if only coordinate $x_1(t)$ is recorded for system $X$, one can construct the vectors
$ \widetilde{x}_1(t) = \left(x_1(t),\,x_1(t-\tau_e),\,x_1(t-2\tau_e),...,\,x_1(t-(E-1)\tau_e) \right)\, $
and the projection of the trajectory in this space is guaranteed to be topologically equivalent to the original orbit for almost any choice of the embedding time $\tau_e$, provided that the embedding dimension $E$ is at least twice larger than the fractal dimension of the original attractor.
We highlight that the smallest embedding dimension accomplishing this task is typically smaller in practice: for example, it is well known that the Lorenz attractor can be embedded in a shadow manifold with $E=3$, while the Takens' theorem would require $E \geq 5$ \cite{kennel1992determining}.
Even if this mapping is strictly valid only under the assumption of noise-free measurements, it has been empirically demonstrated to be useful also in the analysis of real-world data, which are unavoidably affected by different sources of noise \cite{sugihara2012detecting, krakovska2022state}.
Consistently with our assumptions, the unobserved dynamic variables of each system cannot be unobserved common drivers, as we postulated that neither $X$ nor $Y$ include autonomous subsets of variables.
The robustness of our approach with respect to to the choice of $E$, $\tau_e$ and the other relevant hyperparameters ($k$ in Eq. \ref{eq:info_imbalance} and $\tau$ in Eq. \ref{eq:ineq}) is discussed in the \emph{Materials and Methods}.

\begin{figure*}[hb!]
\centering
\includegraphics[width=15.5cm]{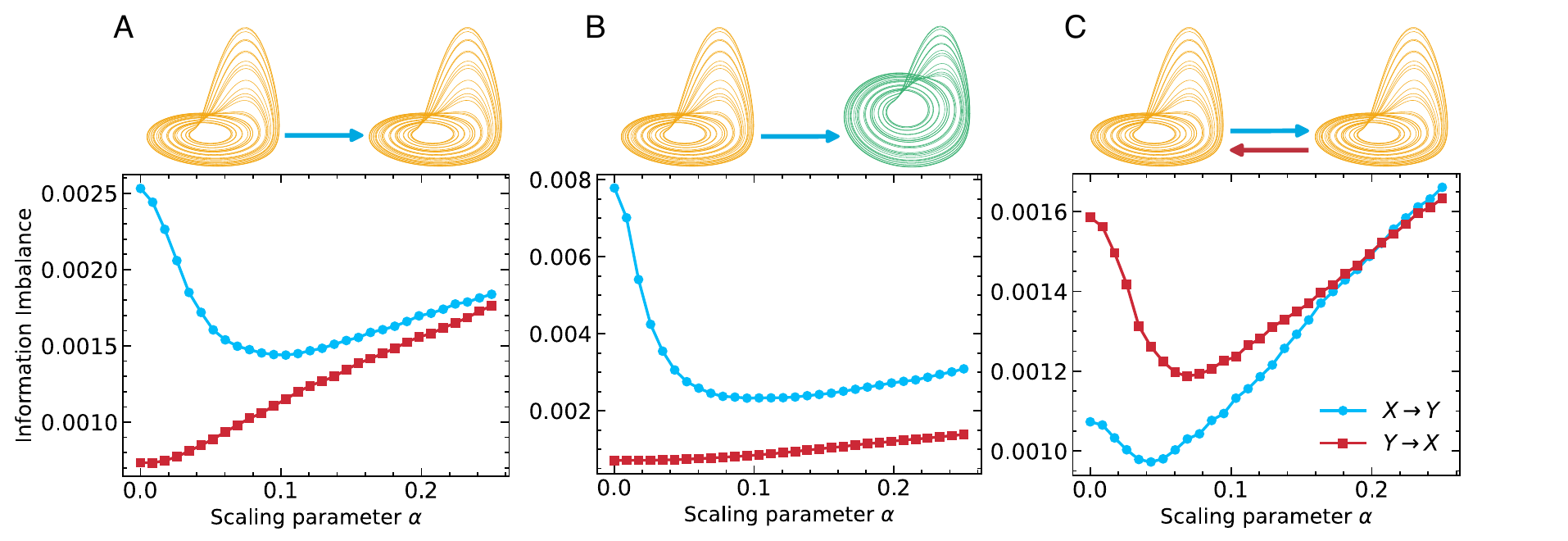}
\caption{Profiles of the Information Imbalance $\Delta(\alpha)$ as a function of $\alpha$, to assess the presence of the causal links $Y\rightarrow X$ and $X\rightarrow Y$. The three panels refer to different pairs of R\"{o}ssler systems: (\emph{A}) identical and unidirectionally coupled with coupling strength $\varepsilon = 0.1293$, (\emph{B}) different and unidirectionally coupled with $\varepsilon = 0.1293$, (\emph{C}) identical and bidirectionally coupled with $\varepsilon_{X\rightarrow Y} = 0.0603$ and $\varepsilon_{Y\rightarrow X} = 0.1$.}
\label{fig:alpha_minimum}
\end{figure*}

\section*{Results}
We first apply our method to model dynamical systems in which a ground-truth causal relationship is defined, and then we validate it on real-world time series using an EEG data set collected in our laboratories.

\subsection*{Causality detection in model systems}
The dynamical systems employed in the following analysis are based on a first-order dynamics that can be generally written as
\begin{subequations}
\begin{align}
    \dot{X} &= f\left( X \right), \label{eq:unidirectional_systems_X} \\
    \dot{Y} &= g\left( Y \right) + \varepsilon\,G\left(X,\,Y \right), \label{eq:unidirectional_systems_Y}
\end{align}
\end{subequations}
in the case of unidirectional coupling ($X\rightarrow Y$), and 
\begin{subequations}
\begin{align}
    \dot{X} &= f\left( X \right) + \varepsilon_{Y\rightarrow X}\,F\left(X,\,Y\right), \\
    \dot{Y} &= g\left( Y \right) + \varepsilon_{Y\rightarrow X}\,G\left(X,\,Y \right),
\end{align}
\end{subequations}
in the case of bidirectional coupling ($X\leftrightarrow Y$).
From a conceptual point of view, we highlight that the word ``causality'' is not used improperly in this context, as these systems satisfy the counterfactual definition that is embraced by modern causal inference in the form of intervention, both in Rubin's potential outcome framework \cite{imbens_rubin_2015} and in Pearl's graphical approach \cite{pearl2009causality}. 
For example, introducting an external forcing in Eq. (\ref{eq:unidirectional_systems_X}) would have a clear effect on the dynamics of $Y$, while disturbing the motion of $Y$ by directly intervening on Eq. (\ref{eq:unidirectional_systems_Y}) would let the motion of $X$ unperturbed.
However, as outlined in the previous Section, our operative approach is based on a predictability principle and does not require the use of external interventions.

In the unidirectional setting we tested two R\"{o}ssler systems \cite{rossler1976anequation} both with identical and with different frequencies, and two 40-dimensional Lorenz 96 systems \cite{lorenz2006predictability} with different forcing constants.
We tested the bidirectional scenario using two identical R\"ossler sytems and two identical Lorenz systems \cite{lorenz1963deterministic}.
All the systems display a chaotic dynamics.
We refer to the \emph{Materials and Methods} for the explicit equations of each pair of systems.
All the tests were performed by extracting from the trajectories $N=5000$ realizations, except for the Convergent Cross Mapping method that requires to monitor the convergence of the results as a function of the number of samples.

In order to benchmark our procedure, we first studied the qualitative behavior of the Information Imbalance $\Delta(\alpha)$ in the optimal scenario where all the coordinates of each system are known, so that the use of time-delay embeddings is unnecessary. 
We employed R\"{o}ssler systems and considered three different coupling configurations, where the link is unidirectional between identical systems (Fig. \ref{fig:alpha_minimum}\emph{A}), unidirectional between different systems (Fig. \ref{fig:alpha_minimum}\emph{B}) and bidirectional between identical systems (Fig. \ref{fig:alpha_minimum}\emph{C}).
In these illustrative examples the number of neighbors $k$ was set to 1 and the time lag $\tau$ was fixed to 5.
In the case of unidirectional coupling (Fig. \ref{fig:alpha_minimum}\emph{A} and \emph{B}), $\Delta(\alpha)$ monotonically increases in the direction where the causal link is absent ($Y\rightarrow X$), while it clearly shows the presence of a minimum in the correct coupling direction ($X\rightarrow Y$). 
Consistently with this scenario, if the coupling is bidirectional the Information Imbalance shows clear minima as a function of $\alpha$ in both directions (Fig. \ref{fig:alpha_minimum}\emph{C}). As a consequence, the Imbalance Gain is positive in both directions.

\begin{figure*}[ht!]
\centering
\includegraphics[width=17.5cm]{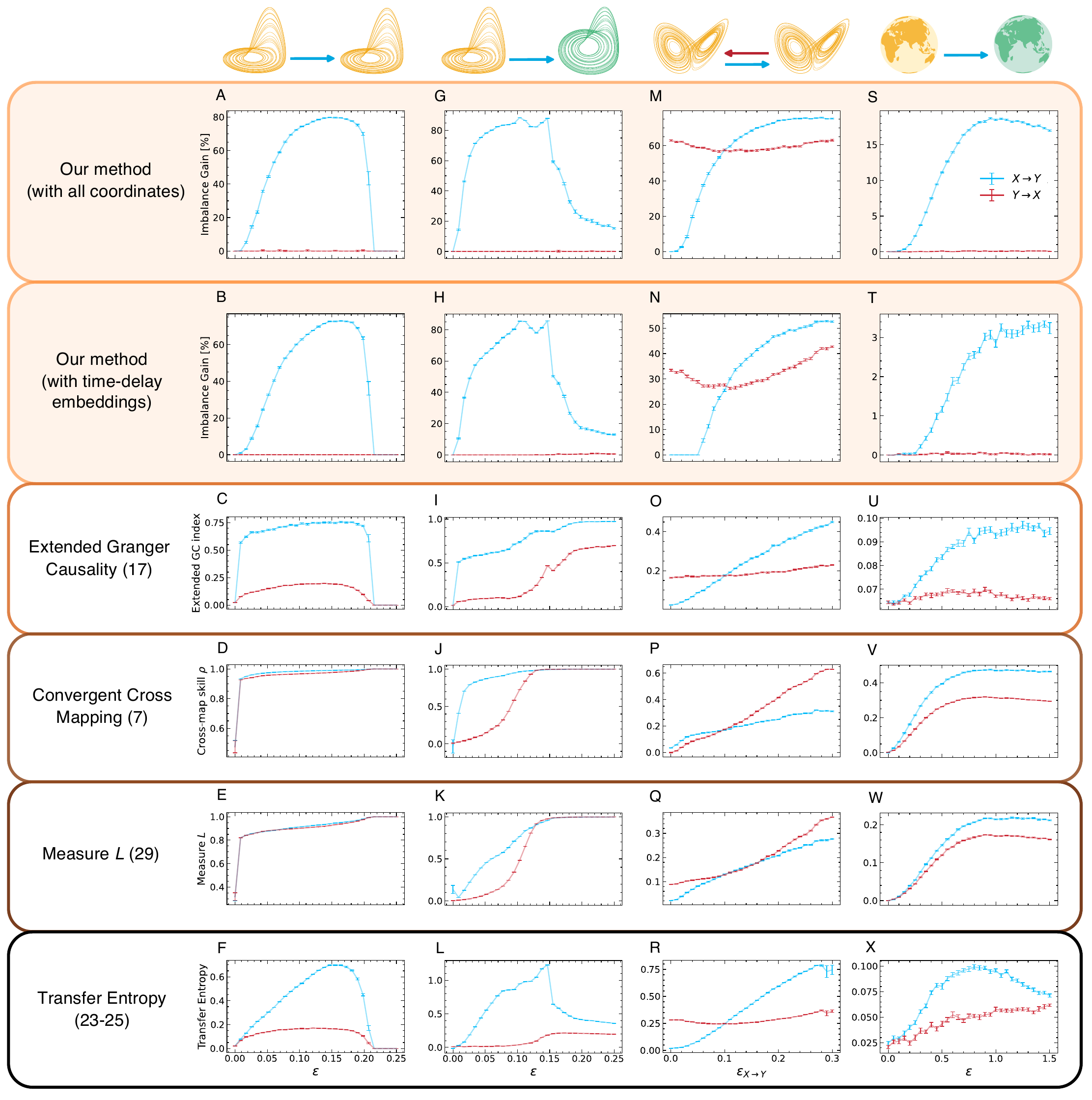}
\caption{Comparison of different causality detection methods. 
The results are shown as a function of the coupling parameter $\varepsilon$ ($\varepsilon_{X\rightarrow Y}$ in the bidirectional case). 
(\emph{A})-(\emph{F}) Identical and unidirectionally coupled R\"{o}ssler systems.
(\emph{G})-(\emph{L}) Different and unidirectionally coupled R\"{o}ssler systems. (\emph{M})-(\emph{R}) Bidirectionally coupled Lorenz systems. 
The coupling $\varepsilon_{Y\rightarrow X}$ was fixed to 0.1. (\emph{S})-(\emph{X}) Unidirectionally coupled Lorenz 96 systems, with 40 variables each ($F_X = 5$, $F_Y = 6$). 
In our method we fixed the embedding time $\tau_e$ to 1 and we employed as embedding dimensions (\emph{B}),(\emph{H}),(\emph{N}) $E = 3$, and (\emph{T}) $E = 30$. The time lag was fixed to $\tau = 20$ for the R\"{o}ssler systems, $\tau = 5$ for the Lorenz systems and $\tau = 30$ for the Lorenz 96 systems.
}
\label{fig:comparison}
\end{figure*}

\renewcommand{\arraystretch}{1.2} 
\begin{table*}[hb!]
\caption{\label{tab:false_positive} False positive rates (FPR), approximated to the first decimal digit, for detections of couplings $Y\rightarrow X$ in systems coupled in direction $X \rightarrow Y$. To exclude the results in the synchronization regime, only couplings configurations with $\varepsilon < 0.18$ were considered for the identical R\"{o}ssler systems, and with $\varepsilon < 0.13$ for the different R\"{o}ssler systems.}
\tabcolsep=0pt
\begin{tabular*}{\textwidth}{@{\extracolsep{\fill}}lcccc@{\extracolsep{\fill}}}
& Identical R\"{o}ssler systems & Different R\"{o}ssler systems & Lorenz 96 systems \\ \hline
 Our method (all coordinates) & 0\% (0/21) & 0\% (0/16) & 9.7\% (3/31) \\
 Our method (embeddings) & 0\% (0/21) & 0\% (0/16) & 12.9\% (4/31) \\
 Extended Granger Causality \cite{Chen2004analyzing}  & 100\% (21/21) & 100\% (16/16) & 100\% (31/31) \\ 
 Convergent Cross Mapping \cite{sugihara2012detecting} & 100\% (21/21) & 93.8\% (15/16) & 96.8\% (30/31) \\
 Measure $L$ \cite{chicharro2009reliable} & 100\% (21/21) & 93.8\% (15/16) & 96.8\% (30/31) \\
 Transfer Entropy \cite{schreiber2000measuring,palus2001synchronization,palus2007directionality} & 100\% (21/21) & 100\% (16/16) & 100\% (31/31) \\
 \hline
\end{tabular*}
\vspace{1.2cm}
\end{table*}

To demonstrate the robustness of the procedure more quantitatively, in Fig. \ref{fig:comparison} we report a comparison of our method with four alternative approaches to assess causality between time-dependent variables, namely the Extended Granger Causality \cite{Chen2004analyzing}, Convergent Cross Mapping \cite{sugihara2012detecting}, the Measure $L$ \cite{chicharro2009reliable} and Transfer Entropy \cite{schreiber2000measuring,palus2001synchronization,palus2007directionality}, which are described in the \emph{SI Appendix} (section 2).
The latter three approaches are model-free, like ours, while the former assumes a local autoregressive model.
Each method produces an estimate for each coupling direction.
As the other methods employ time-delay embeddings, to ensure a fair comparison we also applied our approach using the delayed representations of single coordinates ($x_1$ and $y_1$).
The set of tests was carried out using four different pairs of dynamical systems: two low-dimensional and unidirectionally coupled (identical and different R\"{o}ssler systems, Figs. \ref{fig:comparison}\emph{A} to \emph{F} and Figs. \ref{fig:comparison}\emph{G} to \emph{L}), one low-dimensional and bidirectionally coupled (Lorenz systems, Figs. \ref{fig:comparison}\emph{M} to \emph{R}) and one high-dimensional and unidirectionally coupled (Lorenz 96 systems, Figs. \ref{fig:comparison}\emph{S} to \emph{X}).
To evaluate the statistical significance of the results in a consistent way among the different measures, each point in the panels of Fig. \ref{fig:comparison} was computed as the average of 20 independent estimates, and associated to its standard error. 
The statistical significance of a single Imbalance Gain estimate is assessed with a permutation test over the indices of the putative driver realizations, which generates a null distribution under the  hypothesis of absence of causality (\emph{SI Appendix}, section 6 and Table S1).

In the unidirectionally coupled R\"{o}ssler systems (Figs. \ref{fig:comparison}, first column) and Lorenz systems  (Figs. \ref{fig:comparison}, second column) our method successfully finds a unidirectional link $X\rightarrow Y$, displaying absent or negligible signal in the $Y\rightarrow X$ direction.
The sharp collapse  observed in Fig. \ref{fig:comparison}\emph{A}-\emph{E} at $\epsilon \sim 0.2$ occur in correspondence of the \emph{complete synchronization} of the two systems \cite{boccaletti2002thesync, springer2023data}, where the trajectories of $X$ and $Y$ become identical.
The other methods correctly detect that causality is stronger in the $X\rightarrow Y$ direction than in the reverse one; however, they do not allow deducing from the data that the coupling $Y\rightarrow X$ is absent.

In the bidirectional case (Figs. \ref{fig:comparison}, third column) all the methods correctly detect the presence of both the causal links, but the cross mapping methods do not predict the correct ranking of the two coupling strengths for $\varepsilon_{X\rightarrow Y} \gtrsim 0.1$.
Consistently, in the other methods (Imbalance Gain, Extended Granger Causality and Transfer Entropy) the curves quantifying the strengths of the two causal links intersect at $\varepsilon_{X\rightarrow Y} \simeq 0.1$, which is the value at which the opposite coupling parameter $\varepsilon_{Y\rightarrow X}$ was fixed.
In the high-dimensional scenario (Fig. \ref{fig:comparison}, fourth column), all the approaches detect the correct order of the causal coupling but, once again, the three metrics used for comparison do not allow concluding that causality is actually absent in one direction. 

Table \ref{tab:false_positive} reports the number of false positive detections in the scenarios where the directional coupling $Y\rightarrow X$ is absent, rejecting the null hypothesis of a causal measure being different from zero according to a one-tailed t statistics threshold of $t_{19} = 3.579$ ($p < 0.001$). We report in the \emph{SI Appendix} (section 7 and Fig. S4\emph{A}) an extended description of the statistical test and the false positive rates for other choices of the significance threshold.
The other approaches display a false positive rate close to 100\%. With our measure the false positives are absent in the R\"{o}ssler systems, while they are around 10-13\% in the 40-dimensional Lorenz 96 systems. 
The abrupt reduction in false positive detection is a major advantage of our approach.

\paragraph{Causality detection on EEG time series.}
We employed EEG data to validate our approach in a real-world scenario.
We performed a psychophysical experiment to assess whether the Imbalance Gain could establish the presence of a causal relationship between the experimental manipulation and EEG activity across participants, 
and to understand whether it could be used to study the information flow between different EEG channels.
In the experiment 19 healthy volunteers were asked to judge the duration of two stimuli, a visual and an auditory one, presented sequentially.
Participants' task was to report, by pressing a key, which of the two stimuli was displayed for longer time.
The first stimulus in the pair, the comparison stimulus, was a visual grating varying randomly in display time in each experimental trial.
Seven different durations of the visual stimulus were tested, evenly spaced between 0.3 and 0.9 seconds.
The second stimulus, called standard, was a burst of white noise presented through headphones for 0.6 seconds in each trial. 
The details of the experiment and additional validation tests are reported in the \emph{SI Appendix} (section 8 and Fig. S5).

\begin{figure}[b!]
\centering
\includegraphics[width=8.5cm]{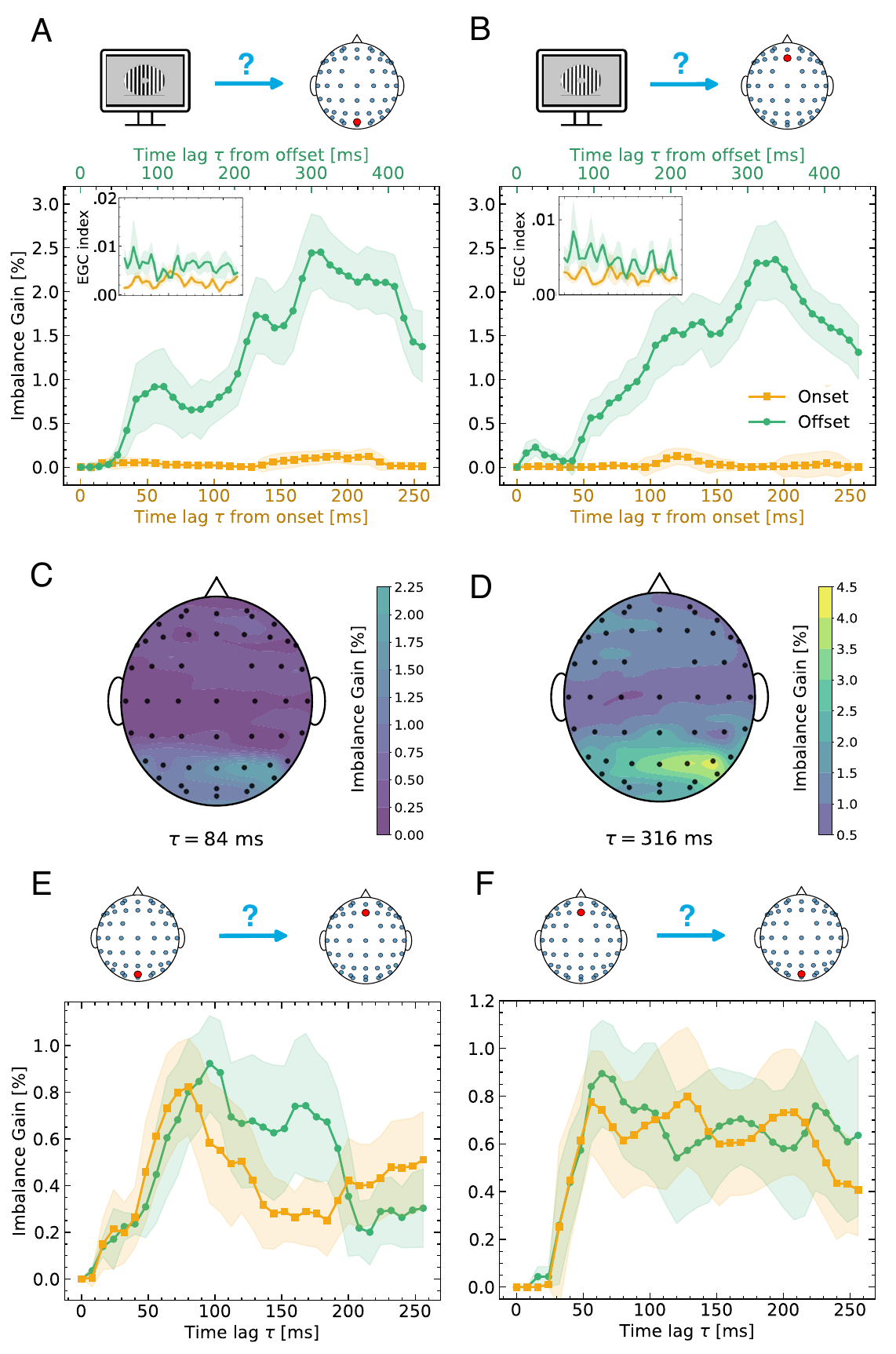}
\caption{(\emph{A})-(\emph{B}): Imbalance Gains to assess the presence of the link from the duration of the stimulus to the EEG signals in two example channels (\emph{A}) POz and (\emph{B}) Fz, as a function of the time lag $\tau$.
The initial time $t=0$ was set to the onset of the stimulus for the orange curves, and to its offset for the green ones.
The shaded area represents the standard error associated to the average Imbalance Gain, computed over 19 participants.
The inset panels display the average profiles of the Extended Granger Causality index for the same tests and time intervals.
(\emph{C})-(\emph{D}): Topoplots displaying the distributions of the Imbalance Gains from all the 51 channels, at two different time lags from the stimulus offset.
(\emph{E})-(\emph{F}): Imbalance Gains for testing the couplings POz $\rightarrow$ Fz and Fz $\rightarrow$ POz, respectively.}
\label{fig:fig_EEG}
\end{figure}

First we investigated the causal link between the duration of the comparison stimulus, which is a categorical variable assuming a value between 0.3 and 0.9 seconds, and the EEG traces relative to its onset and its offset.
Notably, in this application the putative causal variable $X(0)$ appearing in Eq. (\ref{eq:ineq}) is one-dimensional and time-independent.
We excluded from our analysis the information on the second stimulus, as it was shown for a fixed duration among different trials.
The Imbalance Gain was computed independently for each participant using the different trials as independent realizations and employing time-delay embeddings of 44 ms ($E = 12$, $\tau_e = 4$ ms).
The results are shown for two example channels: one parieto-occipital (POz, Fig. \ref{fig:fig_EEG}\emph{A}) and one frontal (Fz, \ref{fig:fig_EEG}\emph{B}).

\noindent We carried out two sets of measurements for different choices of the initial time $t=0$ appearing in Eq. (\ref{eq:ineq}), which corresponds here to the first point of the predictive delay embedding.
In the first tests we set $t=0$ to the stimulus onset and we studied the behavior of the Imbalance Gain $\Delta(\alpha)$ as a function of the time lag $\tau$, limiting $\tau$ to a window before the offset of the shortest stimulus.
In this setting, trials corresponding to different durations of the stimulus cannot be yet differentiated because the different comparison stimuli are all indistinguishable in the time window considered; as a consequence, no coupling duration $\rightarrow$ EEG activity should be observed.
Consistently with this observation, we detected only a negligible Imbalance Gain within the first 256 ms from stimulus onset (orange traces in Figs. \ref{fig:fig_EEG}\emph{A} and \emph{B}).
Indeed, using a (one-tailed) $t$ statistics threshold of $t_{18} = 3.610$ to reject the null hypothesis of the Imbalance Gain being equal to zero ($p < 0.001$), we found a total false positive rate of $\sim0.5$\% across the channels.
We report the false positive rate as a function of the significance threshold in the \emph{SI Appendix} (Fig. S4\emph{B}).
In the second set of tests we set the initial time $t=0$ to the stimulus offset and we studied the Imbalance Gain within a window of 456 ms.
From a perceptual stand point, this temporal window represents the period in which a signature of the subjective experience of stimulus duration should arise \cite{kononowicz2014decoupling}, as only after the offset of the stimulus its duration information becomes available to the participants.
Therefore, a possible causal influence between the duration of the stimulus and the EEG activity may emerge in this second scenario.
Using the same statistical procedure described above, we detected significant couplings with a different time modulation depending on the channel, whereby the Imbalance Gain started to raise at early latencies in occipital and parietal channels (see Fig. \ref{fig:fig_EEG}\emph{A} and \emph{C}) and peaked in a time span ranging from 300 to 400 ms after the stimulus offset also in frontal channels (see Fig. \ref{fig:fig_EEG}\emph{B} and \emph{D}). 
This result is in agreement with recent EEG studies in the field of time perception which show that within similar latencies, and particularly in fronto-central electrodes, EEG activity contains information about participants' perceived stimulus duration \cite{ofir2022neural,damsma2021temporal}. 

As a comparison, we applied the Extended Granger Causality approach to the same causal detection task.
We tested different combinations of $E$ and $k$ in order to maximize the difference of the signals in the offset and onset periods, and to minimize at the same time the rate of false positives in the second scenario (see section 2B in the SI Appendix for details).
In the optimal case (insets of Figs. \ref{fig:fig_EEG}\emph{A} and \emph{B}) we could observe only a faint signal after stimulus offset, and a total false positive rate of $\sim43$\% after stimulus onset, around 2 orders of magnitude more than with our approach.

Our method is able to retrieve a signature of participants' perceptual decision-making processes and its link to task performance. To illustrate this we studied the causal relationships between two extremes of an hypothetical brain network functionally related to duration processing. 
Specifically, we evaluated the causal link between a parieto-occipital electrode, POz, and a fronto-central one, Fz.
The activity of the former is supposedly linked with an early stage of duration processing \cite{tonoyan2022time} where stimulus sensory and duration information is encoded and conveyed to downstream brain regions (duration encoding), while the latter is associated to an higher-level processing stage where duration information is read out and used to perform the task at hand (duration decoding) \cite{Protopapa2019,Protopapa2023}.
We computed the Imbalance Gain POz $\rightarrow$ Fz within 256 ms from stimulus' comparison onset and offset (yellow and green traces in Fig. \ref{fig:fig_EEG}\emph{E}, using time-delay embeddings of 44 ms for both the signals.
We found that the Imbalance Gain relative to the different periods (i.e., onset or offset) changed differently as function of $\tau$ (period-tau interaction: $F_{26,468} = 1.529, p = 0.04$).
In both periods the Imbalance Gain peaked around 90 ms, time lag which may reflect the propagation delay in the information flow between the two channels.
However, in the onset period the signal slowly decayed until it reached a plateau, whereas after stimulus offset we observed a second peak in Imbalance Gain around $\tau$ = 160 ms.
Interestingly, previous works have shown that, in the same electrode (Fz) and within similar latencies ($\sim$ 150 ms), it is possible to detect decision-related EEG activity which originates from feed-forward communication from the visual cortices \cite{Thorpe1996, Fabre-Thorpe2001}.
No such effect of the interaction between $\tau$ and period on the Imbalance Gain was found in the case Fz $\rightarrow$ POz ($F_{26,468} = 0.448$, see Fig. \ref{fig:fig_EEG}\emph{F}).
To better characterize the relationship between our results and participants' performance we computed the Kendall's correlation coefficient between the Imbalance Gain and participants' accuracy in the task.
The results of this analysis show that a positive association is present only at the offset period in correspondence to the Imbalance Gain peaks (in particular around 160 ms, see \emph{SI Appendix}, Fig. S5). Although more experiments are needed to better characterize this result, this shows that the Imbalance Gain is a promising measure to investigate the link between brain functional connectivity and behavior.

\section*{Discussion}
We proposed an approach to detect causality in time-ordered data based on the Information Imbalance, a statistical measure constructed on distance ranks.
The underlying idea is to quantify how the description of a subset of variables in the future is affected by the addition of the putative causal variables in the past; in this sense, our method can be seen as a nonlinear and model-free generalization of Granger causality, 
suitable for high-dimensional data.

Our approach is also related with the measure $L$ \cite{chicharro2009reliable}, which, using our notation, is based on the comparison of $\Delta\left(d_{X(0)}\rightarrow d_{Y(0)}\right)$ with $\Delta\left(d_{Y(0)}\rightarrow d_{X(0)}\right)$ (\emph{SI Appendix}, section 2A). This simpler approach faces the limitations illustrated in Fig. \ref{fig:comparison}: the evaluation of a single inequality only allows to identify the dominant causal link, without recognizing situations where the coupling in one direction is absent.


In case of missing dynamic variables, our measure can be applied with time-delay embeddings.
In principle, Takens' theorem states that the time-delay embeddings of a single coordinate of $Y$, e.g. $\widetilde{y}_1$, should allow to reconstruct the full system $(X,Y)$ even when only the coupling $X\rightarrow Y$ is present.
If this was the case, even when $X$ causes $Y$ there would be no $\alpha > 0$ satisfying Eq. (\ref{eq:ineq}) if the distance spaces were constructed with time-delay embeddings, because any information carried by $X$ would be redundant.
However, the above statement is correct only if noise is absent, and if it is possible to record the coordinates with arbitrary precision \cite{kennel1992determining}.
In practical applications, where the observation time is finite and measurement/integration noise is present, the best reconstruction carries only a \emph{partial} information of $X$, so that $\widetilde{x}_1$ actually contains unique information about the future of $\widetilde{y}_1$. This scenario is supported by the numerical experiments presented in this work.

By testing our method on several coupled dynamical systems with different coupling configurations, we found that
our measure is significantly more robust than the compared methods against the drawback of false positives. 
In low-dimensional systems the difference in performance with other approaches is striking (see Table \ref{tab:false_positive}). In the high-dimensional scenario, some false positive detections are present, but comparing the signals in the two directions allows to clearly discern the irrelevance of one of the two couplings (see Fig. \ref{fig:comparison}, panels P and Q).
The advantage over the other approaches is still evident when the dynamics of driver and driven systems involve different time scales (\emph{SI Appendix}, section 9 and Fig. S7).


We further applied our measure to real-world time series from EEG experiments, studying the time modulation of causal links from an experimental and static variable - the duration of the comparison stimulus - to the channel-dependent EEG activity, and between EEG activities of different channels.
In the tests uncovering causal signals, the time modulations of the Imbalance Gain are consistent with the latencies at which the information of stimulus duration is expected to be transferred from the visual cortex to the fronto-central channels and elaborated by the latters \cite{ofir2022neural,damsma2021temporal}.
Our findings suggest that applying our approach to \emph{ad hoc} experiments may provide new insights into functional and effective brain connectivity.


We introduced our method within a framework involving two systems $X$ and $Y$, without considering the potential influence of a third observed system, $Z$.
In the \emph{SI Appendix} (section 3) we show that our approach can be generalized to include such a third system, and how the causal analysis is affected when $Z$ is a common driver ($X\,\leftarrow\,Z\,\rightarrow\,Y$) or an intermediate system (e.g. $X\,\rightarrow\,Z\,\rightarrow\,Y$) \cite{leng2020partialCM}.

In conclusion, we believe that the benchmark presented in this work demonstrate that this approach overcomes relevant limitations of other model-free causality detection methods.
Its stronger statistical power shows up in particular when applied to high-dimensional systems in which \emph{causality is absent}, a scenario which is of the utmost important for understanding how to control a system.
This, we believe, will trigger the interest of scientists working on causality detection with real-world time-dependent data.

\section*{Materials and Methods}
\subsection*{\label{appendix:dynamical_systems}Details on the dynamical systems}
  
All the analysis on the dynamical systems reported in the Results Section were carried out discarding the first 100000 samples of the generated time series.
All the dynamical systems were integrated using the 8-th order explicit Runge-Kutta method DOP853 in the Python library SciPy, except for the coupled Lorenz 96 systems for which the SciPy implementation of the LSODA integrator was employed \cite{virtanen2020SciPy}.

\subsubsection*{R\"{o}ssler systems}
In the unidirectional case $X\rightarrow Y$, the equations of the coupled R\"{o}ssler systems are
\begin{subequations}
\begin{align}
    &\dot{x}_{1}(t)=-\omega_{1}\, x_{2}(t)-x_{3}(t) \label{eq:eq_x1_rosslers}\\
    &\dot{x}_{2}(t)=\omega_{1}\, x_{1}(t)+0.15\, x_{2}(t) \\
    &\dot{x}_{3}(t)=0.2+x_{3}(t)\left[x_{1}(t)-10\right] \\
    &\dot{y}_{1}(t)=-\omega_{2}\, y_{2}(t)-y_{3}(t) + \varepsilon\left(x_1(t) - y_1(t)\right) \\
    &\dot{y}_{2}(t)=\omega_{2}\, y_{1}(t)+0.15\, y_{2}(t) \\
    &\dot{y}_{3}(t)=0.2+y_{3}(t)\left[y_{1}(t)-10\right]\,,
\end{align}
\end{subequations}
with $\omega_1 = \omega_2 = 1.015$ for the case of identical systems, and $\omega_1 = 1.015$, $\omega_2 = 0.985$ for the case of different systems, as studied in Refs. \cite{palus2007directionality,palus2018causality}. 
The term $\varepsilon_{Y\rightarrow X}\left(y_1(t)-x_1(t)\right)$ was added to Eq. (\ref{eq:eq_x1_rosslers}) in the bidirectional tests.
Both the initial state of system $X$ and the initial state of $Y$ were set by multiplying the components of the vector (10, 10, 10) by three random numbers between 0.5 and 1.5.
For all the tested values of the coupling constants, the equations were integrated with a fixed time step of 0.0785 and the time series was downsampled with a frequency of $1/4$, resulting in a sampling step of 0.314. After discarding the first 100000 samples, 105000 were saved and employed in the analysis.

\subsubsection*{Lorenz systems}
The bidirectionally coupled Lorenz systems are defined by the following equations:
\begin{subequations}
\begin{align}
    \dot{x}_{1}(t) &=10\,(x_2(t)-x_1(t)) \\
    \dot{x}_{2}(t) &=x_1(t)(28-x_3(t)) - x_2(t)  + \varepsilon_{Y\rightarrow X}\,    y_1^2(t) \\ 
    \dot{x}_{3}(t) &= x_1(t)\,x_2(t) - 8/3\, x_3(t) \\
    \dot{y}_{1}(t) &=10\,(y_2(t)-y_1(t)) \\
    \dot{y}_{2}(t) &=y_1(t)(28-y_3(t)) - y_2(t)  + \varepsilon_{X\rightarrow Y}\,    x_1^2(t) \\ 
    \dot{y}_{3}(t) &= y_1(t)\,y_2(t) - 8/3\, y_3(t)
\end{align}
\end{subequations}
The equations were integrated using the time step $dt = 0.01$, initializing the system with the same protocol used for the R\"{o}ssler systems. The resulting time series were saved with a downsampling frequency of $1/5$, resulting in a sampling step of $0.05$ and $205000$ samples.
    
\subsubsection*{Lorenz 96 systems}
Using the conventions $x_{-1}=x_{N-1},\,x_0=x_N$ and $x_{N+1}=x_1$, the equations of the 40-dimensional unidirectionally coupled Lorenz 96 systems employed in the tests are
\begin{subequations}
\begin{align}
    \dot{x}_i &= (x_{i+1}-x_{i-2})\,x_{i-1} - x_i + F_X\,, \\
    \dot{y}_i &= (y_{i+1}-y_{i-2})\,y_{i-1} - y_i + F_Y + \varepsilon\,x_i\,,
\end{align}
\end{subequations}
with $i=1,...,40$. The initial state was set to $x_i(0) = F_X$ ($y_i(0) = F_Y$) for $i>1$ and to $x_1(0) = F_X + \mathcal{R}$ ($y_1(0) = F_Y + \mathcal{R'}$) for the first component, where $\mathcal{R}$ ($\mathcal{R'}$) is a random number between 0 and 1.
The equations were integrated with the time step $dt = 0.03$ and the trajectories were downsampled with frequency $1/2$. 
The tests reported in Fig. \ref{fig:comparison} were carried out using trajectories with 252500 points, while the analysis reported in Fig. \ref{fig:fig_emb} were carried out over realizations with 752000 samples.

\subsection*{Average Imbalance Gain}
In Figs. \ref{fig:comparison}, \ref{fig:fig_EEG} and \ref{fig:fig_emb}, where we show an \emph{average} Imbalance Gain, we inferred one optimal scaling parameter $\alpha$ for all the independent estimates.
The optimal parameter was obtained by maximizing the $\alpha$-dependent Imbalance Gain $\langle \left(\Delta(\alpha=0) - \Delta(\alpha)\right) / \Delta(\alpha=0) \rangle$, where the brackets identify the average across different estimates.

\subsection*{Robustness with respect to the hyperparameters}

\begin{figure}
\centering
\includegraphics[width=8cm]{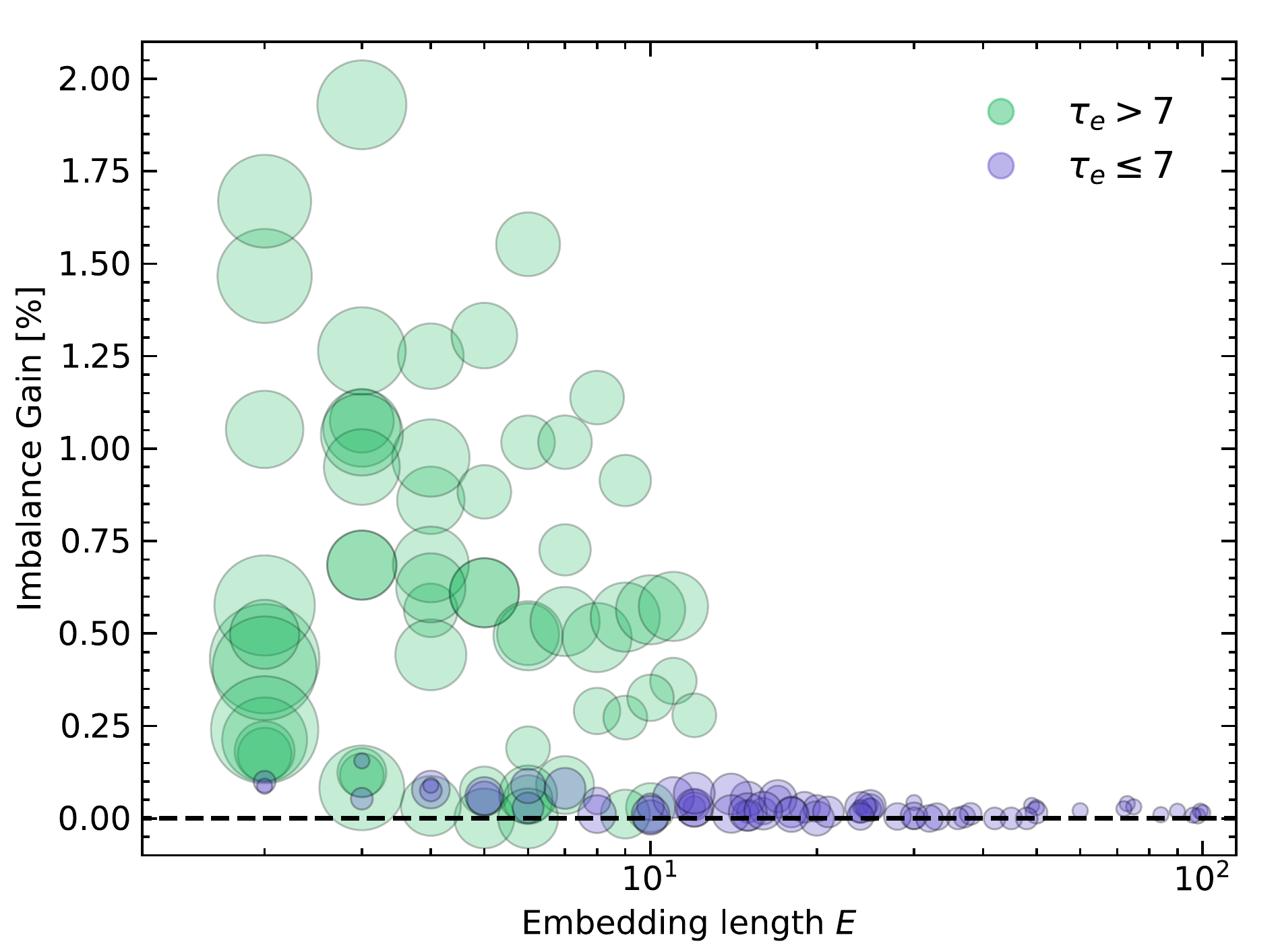}
\caption{Imbalance Gain as a function of the embedding length $E$, computed in direction $Y\rightarrow X$ for two Lorenz 96 systems with opposite coupling $X\rightarrow Y$ ($F_X = 6$, $F_Y = 4$, $\varepsilon = 1$), using $\tau = 30$. The areas of the circles are proportional to the embedding times $\tau_e$ employed in the reconstruction. Each point is the average over 20 independent estimates.
}
\label{fig:fig_emb}
\end{figure}

We investigated how the Imbalance Gain is affected by the embedding parameters $E$ and $\tau_e$ on a pair of Lorenz 96 systems with $F_X = 6$ and $F_Y = 4$, coupled in direction $X\rightarrow Y$.
In Fig. \ref{fig:fig_emb} we show the Imbalance Gain in a case in which it should be equal to zero (absence of causality) for different choices of $E$ (along the $x$-axis) and $\tau_e$ (proportional to the area of the circles), restricting to a maximum window length $(E-1)\,\tau_e = 200$.
Due to the absence of a causal link all the points should ideally lie on the dashed line (zero Imbalance Gain).
Our measure provides false positive detections for $E\lesssim 10$, which are particularly evident when the embedding time $\tau_e$ is too large ($\tau_e > 7$, green circles).
On the other hand, when $E$ is sufficiently large and $\tau_e$ is sufficiently small, the Imbalance Gain appears robust against the detection of false positives for several different choices of the embedding parameters ($\tau_e \leq 7$, violet circles).
In practical applications we suggest to fix $\tau_e$ to a small value (of the same order of the sampling time) and compute $\delta\Delta$ in a scenario without causality for increasing values of the embedding dimension $E$, considering the result reliable if $\delta\Delta$ converges to zero.
All the analysis shown in this work were performed following this criterion.

Another hyperparameter of our approach is $k$, the number of neighbors used to compute the Imbalance Gain (see eq. (\ref{eq:info_imbalance})). A large value of $k$ reduces the statistical uncertainty but can bias the estimate towards the absent coupling scenario, as the Imbalance Gain is deterministically equal to 0 in the limit case $k = N-1$. As a rule of thumb we suggest a value of $k$ of at most 5 \% of the available data. Choosing $k$ in a wide range of values consistent with this choice ($k\sim 10-250$ for the Lorenz 96 systems with $N = 5000$) does not affect the final outcome of the causal analysis in the examples we considered (\emph{SI Appendix}, Fig. S8). All the results in this work were obtained using $k=1$ for the three-dimensional dynamical systems, and setting $k=20$ for the Lorenz 96 systems and the EEG analysis.

The last hyperparameter is the value of $\tau$, the delay between the time of observation and the time of the prediction. When the only purpose of the test is assessing the presence or absence of the causal link, the Imbalance Gain provides stable results against different choices of $\tau$ (\emph{SI Appendix}, Fig. S9). On the other hand, as shown in the analysis of EEG data, a systematic scan of different values of $\tau$ can provide additional insights into the information transfer between driver and driven variables.
However, we underline that the value of $\tau$ for which the Imbalance Gain displays a maximum can only be interpreted as the time lag at which a perturbation in the driven system $X$ becomes observable in the driven system $Y$, and hence the state of the former becomes (maximally) predictive with respect to the state of the latter.
This property depends in general both on the actual coupling delay in the interaction $X\rightarrow Y$ and on the Lyapunov exponents of the driven system $Y$. Therefore, the maximum of the Imbalance Gain as a function of $\tau$ cannot be taken as a measure of the coupling delay.

As any statistical test, the outcome of our approach depends on the number of samples $N$, which is a property of the data set rather than a proper hyperparameter. 
We show in the \emph{SI Appendix} (Fig. S10) that the scaling of the Imbalance Gain measure as a function of $N$ is comparable to the behaviors of both Measure $L$ and Transfer Entropy, with the advantage of a stable and negligible signal in the direction without coupling in the small $N$ regime.

\subsection*{EEG analysis}
The experiment was conducted on nineteen healthy volunteers (mean age: 24.10, SD: 3.43, 8 females), na\"{i}ve to the purpose of the experiment, none of them reporting any neurological disease. Participants gave their written informed consent before taking part in the study, and were compensated for their participation with 12 Euro/hour. The study was carried out in accordance with the Declaration of Helsinki, and was approved by the ethics committee of the Scuola Internazionale Superiore di Studi Avanzati (SISSA) protocol number 23970.
Further details on the experiment are reported in the \emph{SI Appendix} (section 8).

Statistical assessment on Imbalance Gain data was performed using SciPy \cite{virtanen2020SciPy} and statsmodels \cite{seabold2010statsmodels} packages in Python.
In particular we performed a repeated measures ANOVA to understand the effect of the time lag $\tau$ and the period (i.e., onset and offset) on the Imbalance Gain in the case POz $\rightarrow$ Fz and Fz $\rightarrow$ POz.
This analysis was performed using the \emph{AnovaRM} function with $\tau$ and period as within-subjects factors, considering only values of $\tau$ larger than 44 ms in order to avoid any overlap between the time-delay embeddings at time 0 and time $\tau$.

\subsection*{\label{appendix:data_codes}Data and code availability}

The EEG data analyzed in this study are available at \url{https://osf.io/6jpvg/}. 
Supporting codes are available at the GitHub repository \url{https://github.com/vdeltatto/imbalance-gain-causality.git}.
The repository also contains the codes to generate the trajectories of the dynamical systems analyzed in this work.
The Information Imbalance measure is also implemented in the Python package DADApy \cite{glielmo2022dadapy}.

\section*{Acknowledgments}
We would like to thank Prof. Ali Hassanali, Prof. Valerio Lucarini, Prof. Antonietta Mira, Prof. Christoph Zechner and Dr. Michele Allegra for providing useful comments and insights, and Dr. Yelena Tonoyan for contributing to EEG data collection and preprocessing.  This work was partially funded by NextGenerationEU through the Italian National Centre for HPC, Big Data, and Quantum Computing (grant number CN00000013).

\printbibliography[keyword={main-text}]

\clearpage
\onecolumn
\resetlinenumber
\setcounter{figure}{0}
\renewcommand{\thefigure}{S\arabic{figure}}
\setcounter{table}{0}
\renewcommand{\thetable}{S\arabic{table}}
\setcounter{page}{0}
\thispagestyle{empty}

\vspace*{80px}
\section*{\hfill Supplementary Information for \hfill}

\vspace{0.3cm}
\Large{\textbf{Robust inference of causality in high-dimensional dynamical processes from the Information Imbalance of distance ranks}}

\normalsize
\vspace{0.3cm}
\noindent Vittorio Del Tatto$^1$, Gianfranco Fortunato$^1$, Domenica Bueti$^1$, and Alessandro Laio$^{1,2}$

\vspace{0.3cm}
\indent $^1$ Scuola Internazionale Superiore di Studi Avanzati, SISSA, via Bonomea 265, 34136 Trieste, Italy

\indent $^2$ ICTP, Strada Costiera 11, 34151 Trieste, Italy

\newpage

\section{\label{appendix:copulas}Information-theoretic formulation of the method}

The bridge between the Information Imbalance measure and information theory rests on the statistical theory of copulas. Given a two-dimensional random variable $\mathbf{R} = (R_A,\,R_B)$ with joint density $p_{AB}(R_A,\,R_B)$ and marginal densities $p_A(R_A),\,p_B(R_B)$, we define its \emph{copula variable} as
\begin{equation}
    \mathbf{C} = (c_A,\,c_B) = \left(F_A(R_A),\, F_B(R_B)\right)\,,
\end{equation}
where $F_A$ is the cumulative distribution function of $R_A$: $F_A(R) = \int_{-\infty}^{R}\text{d}R'\,p_A(R')$ (similarly for $F_B$). 
The relevance of copula variables lies in the two following results:
\begin{itemize}
    \item the marginals of $\mathbf{C}$ are uniformly distributed in $[0,1]$, result known as \emph{probability integral transform} \cite{casella2001statistical_SI}:
    \begin{equation}\label{eq:uniform_distr}
        c_A \sim U_{[0,1]}\,,\hspace{0.3cm} c_B \sim U_{[0,1]}\,;
    \end{equation}
    \item according to Sklar's theorem \cite{nelsen2006anintroduction_SI}, the joint density of $\mathbf{R}$ can be decomposed as the product of the marginal densities and a joint copula distribution:
    \begin{equation}
        p_{AB}(R_A,\,R_B) = \widetilde{p}_{AB}(c_A,\,c_B)\,p_A(R_A)\,p_B(R_B)\,.
    \end{equation}
\end{itemize}
In particular, Sklar's theorem implies that the joint entropy of ($R_A,\,R_B$) can be decomposed as the sum of the marginal entropies and the joint entropy of the copula variables $(c_A,\,c_B)$: 
\begin{equation}
    H(R_A,\,R_B) = H(R_A) + H(R_B) + H(c_A,\,c_B)\,.
\end{equation}
From Eq. (\ref{eq:uniform_distr}) it follows that the marginal entropy of each copula variable is zero; as a consequence, 
\begin{equation}
    H(c_A,\,c_B) = H(c_B \mid c_A) = H(c_A \mid c_B)\,.
\end{equation}
Given these properties, the mutual information between variables $R_A$ and $R_B$ can be expressed equivalently as the negative conditional entropy of their copula variables \cite{calsaverini2009aninformation_SI,safaai2018information_SI}:
\begin{subequations}
    \begin{align}
        I(R_A,\,R_B) &= H(R_A) + H(R_B) - H(R_A,\,R_B) \\
        &= -H(c_A\mid c_B) = -H(c_B\mid c_A) \label{eq:MI_copula_entropy}
    \end{align}
\end{subequations}

\subsection{The Information Imbalance as an upper bound of a mutual information}

Given a data set $\{x_i \}_{i=1}^N$ drawn from the unknown probability density $p(x)$, from hereon we call $R_A$ and $R_B$ the RVs describing pair-wise distances measured with the metrics $d_A$ and $d_B$, respectively, and we denote with $R^{A}_{ij}$ and $R^{B}_{ij}$ their empirical realizations:
\begin{subequations}
\begin{align}
    R^{A}_{ij} &\doteq d_{A}(x_i,x_j)\,,\\
    R^{B}_{ij} &\doteq d_{B}(x_i,x_j)\,.
\end{align}
\end{subequations}
With a change of variable, the probability density of the distances from a point $x_i$ in space $d_A$ can be written as
\begin{equation}
    p(R_A\,|\, x_i) \doteq p(R_A^i) = \int \text{d}\tilde{x}\, p(\tilde{x}) \,\delta\left(d_A(\tilde{x},x_i) - R_A^i \right)\,,
\end{equation}
where the integral is over the support of $p(x)$ and $\delta$ denotes the Dirac delta function. The distribution $p(R_B^i)$ can be defined analogously.
Given these distributions, the relative information content of the two distance spaces conditioned on $x_i$ can be measured
by the mutual information $I(R_A^{\,i}, R_B^{\,i})$, which can be cast into a conditional copula entropy according to Eqs. (\ref{eq:MI_copula_entropy}).
However, if we aim at quantifying how well each distance can predict the other, the mutual information is not a suitable measure.
Indeed, we expect that the descriptive power of distance $d_A$ with respect to distance $d_B$ will change if we invert the roles of the two spaces, while the mutual information is symmetric under the exchange of its arguments.

Guided by the intuition that a ``good'' distance should first and foremost describe points that are ``similar'' as close, in ref. \cite{glielmo2022ranking_SI} we proposed to introduce an asymmetry by defining a \emph{restricted mutual information}:
\begin{equation}\label{eq:restricted_MI}
    I^\varepsilon\left(R_A^i\rightarrow R_B^i \right) \doteq -\int_0^\varepsilon \text{d}\tilde{c}_A^{\,i}\, H(c_B^{\,i}\mid c_A^{\,i} = \tilde{c}_A^{\,i})\,.
\end{equation}
Here $c_A^i$ is the copula variable associated to $R_A^i$ and its empirical values can be simply computed as $\{\frac{1}{N}r_{ij}^A\}$, where $r_{ij}^A$ is the rank of point $j$ relative to point $i$ according to distance $d_A$ (similarly in space $B$). This reflects the fact that distance ranks and copulas are statistically identical variables.
In the limit of small $\varepsilon$, the restricted mutual information of Eq. (\ref{eq:restricted_MI}) quantifies the information shared by the distances $d_A$ and $d_B$, constraining the first to a small neighborhood around $x_i$.
With the assumption that large distances are noninformative, we interpret this measure as the actual information in space $d_B$ which is contained in $d_A$.
If we expand the restricted mutual information in the limit $\varepsilon \rightarrow 0$, the first nonzero term is its first derivative around zero:
\begin{subequations}
\begin{align}
    \lim_{\varepsilon\rightarrow 0} \frac{I^\varepsilon\left(R_A^i\rightarrow R_B^i \right)}{\varepsilon} &= 
    \frac{\partial}{\partial \varepsilon} I^\varepsilon\left(R_A^i\rightarrow R_B^i \right)\rvert_{\varepsilon=0} \\
    &= -\lim_{\varepsilon\rightarrow 0} H(c_B^i\mid c_A^i=\varepsilon) \label{eq:cond_copula_entropy} \\
    &= \lim_{\varepsilon\rightarrow 0}\int_0^1 \text{d}c_B^i\,p(c_B^i\mid c_A^i=\varepsilon) \log p \left( c_B^i\mid c_A^i=\varepsilon\right).
\end{align}
\end{subequations}
In our finite data set, we can translate the condition $c_A^i = \varepsilon$ in the limit of small $\varepsilon$ by taking the first $k$ neighbors of $i$, i.e. the points $j$ such that $c_{ij}^A \leq \frac{k}{N}$.
Since the number $k$ should be much smaller than $N$ to fulfill the continuous condition $\varepsilon \rightarrow 0$, we can only access a small number of samples of the density $ p \left( c_B^i\mid c_A^i=\varepsilon\right)$, and a direct estimate of Eq. (\ref{eq:cond_copula_entropy}) would be unreliable. 
Instead of trying to compute Eq. (\ref{eq:cond_copula_entropy}) explicitly, we suppose to know only the average of $ p \left( c_B^i\mid c_A^i=\varepsilon\right)$, that we estimate in the limit $\varepsilon\rightarrow 0$ as
\begin{equation}\label{eq:average_copula_ranks}
    \langle c_B^i\mid c_A^i = \varepsilon \rangle \simeq 
    \frac{1}{N}\langle r_B^i\mid r_A^i = \varepsilon \rangle
    := \frac{1}{N\,k}\sum_{\underset{r_{ij}^A\leq k}{j}} r_{ij}^B\,,
\end{equation}
and we look for an upper bound of the copula entropy in Eq. (\ref{eq:cond_copula_entropy}), using a maximum entropy approach. 
We first observe that the average in Eq. (\ref{eq:average_copula_ranks}) tends to 0 in the case of $d_A$ being maximally informative with respect to $d_B$, and to 1/2 in the fully non-informative case (namely when $ p \left( c_B^i\mid c_A^i=\varepsilon\right) = p \left( c_B^i\right)$). The Information Imbalance defined in Eq. (1) in the main text is then a rank-based estimate of this quantity, with an additional global average over all the data points:
\begin{equation}\label{eq:info_imbalance_copula_avg}
    \Delta (A\rightarrow B) \simeq 2\,\Big\langle\lim_{\varepsilon\rightarrow 0} \langle c_B^i\mid c_A^i = \varepsilon \rangle \Big\rangle_{i=1,...,N}\,.
\end{equation}

It is well known that among all the probability densities with support $[0,+\infty)$ and assigned mean, the one with maximum entropy is exponential \cite{dowson1973maxent_SI}.
The same can be shown in the case of distributions with support $[0,1]$, by using the technique of Lagrange multipliers. In particular, the probability distribution maximizing the entropy of Eq. (\ref{eq:cond_copula_entropy}) is
\begin{equation}
    p_{\text{ME}}(c_B^i \mid c_A^i = \varepsilon) = \frac{\lambda}{1-e^{-\lambda}}\,e^{-\lambda c_B^i}\,,
\end{equation}
where the parameter $\lambda$ is related to the mean by
\begin{equation}\label{eq:average_copula_lambda}
    \langle c_B^i\mid c_A^i = \varepsilon \rangle = \frac{1}{\lambda} + \frac{1}{1-e^{\lambda}} \,.
\end{equation}
By substitution we find a first upper bound:
\begin{equation}\label{eq:first_upper_bound}
    \lim_{\varepsilon\rightarrow 0} H(c_B^i\mid c_A^i=\varepsilon) \leq -\log \frac{\lambda}{1-e^{-\lambda}} + 1 + \frac{\lambda}{1-e^{\lambda}}\,.
\end{equation}
\noindent The right-hand side of Eq. (\ref{eq:first_upper_bound}) is further maximized by a quantity directly related to the average in Eq. (\ref{eq:average_copula_lambda}):
\begin{equation}\label{eq:second_upper_bound}
    \lim_{\varepsilon\rightarrow 0} H(c_B^i\mid c_A^i=\varepsilon) < \log\left(\lim_{\varepsilon\rightarrow 0}\langle c_B^i\mid c_A^i = \varepsilon \rangle\right) + 1\,.
\end{equation}
In particular, the right-hand sides of Eqs. (\ref{eq:first_upper_bound}) and (\ref{eq:second_upper_bound}) are asymptotically equal in the limit $\lambda \rightarrow 0$, namely when $\langle c_B^i\mid c_A^i = \varepsilon \rangle \rightarrow 0$.
This is the case of $d_A$ being maximally informative with respect to $d_B$.
The last inequality, aside with Eq. (\ref{eq:info_imbalance_copula_avg}), allows to write the Information Imbalance as an upper bound of an information-theoretic measure:
\begin{equation}\label{eq:info_imbalance_inequality}
    \Delta(A\rightarrow B) \gtrsim 2\Big\langle \exp\left( \lim_{\varepsilon\rightarrow 0} H(c_B^i\mid c_A^i=\varepsilon) -1 \right) \Big\rangle_{i=1,...,N}\,.
\end{equation}
Equivalently, in terms of the restricted mutual information, the inequality takes the form
\begin{equation}
    \Delta(A\rightarrow B) \gtrsim 2\Bigg\langle \exp\left( -\lim_{\varepsilon\rightarrow 0} \frac{I^\varepsilon\left(R_A^i\rightarrow R_B^i \right)}{\varepsilon} -1 \right) \Bigg\rangle_{i=1,...,N}\,.
\end{equation}
Due to the asymptotic relationship mentioned above, the Information Imbalance provides a better estimate of the right-hand side of Eq. (\ref{eq:info_imbalance_inequality}) in the informative regime, namely when $\Delta(A\rightarrow B)$ is close to 0.

\subsection{Interpretation of the upper bound} 

A straightforward interpretation of the restricted mutual information introduced in Eq. (\ref{eq:restricted_MI}) in the limit $\varepsilon\rightarrow 0$ can be provided by rewriting Eq. (\ref{eq:cond_copula_entropy}) in terms of the distance variables $R_A^i$ and $R_B^i$.
Applying Sklar's theorem and defining $\eta = F_A^{-1}(\varepsilon)$ we can write
\begin{subequations}
\begin{align}
     \lim_{\varepsilon\rightarrow 0} \frac{I^\varepsilon\left(R_A^i\rightarrow R_B^i \right)}{\varepsilon} &= \lim_{\eta\rightarrow 0}\int_{0}^{+\infty}\text{d}\,R_B^i\,p(R_B^i\mid R_A^i=\eta)\,\log\frac{p(R_B^i\mid R_A^i=\eta)}{p(R_B^i)} \\
    &= \lim_{\eta\rightarrow 0}\,D_{KL}\big[p(R_B^i\mid R_A^i=\eta) \parallel p(R_B^i) \big]\,,\label{eq:KL_divergence}
\end{align}
\end{subequations}
where $D_{KL}$ denotes the Kullback-Leibler divergence, which is a pseudodistance between probability densities.
Eq. (\ref{eq:KL_divergence}) clarifies the concept of shared information among distance spaces: when we condition the distribution of distances in $d_B$ to close points according to $d_A$, the greater the change in the distribution, the more informative $d_A$ is with respect to $d_B$.
This allows to translate in an information-theoretic framework the simple intuition that $d_A$ carries information about $d_B$ when close points in the former remain close in the latter.

If we specialize to the problem of causal detection by taking the distance spaces $d_A=d_{\alpha X(0),Y(0)}$ and $d_B=d_{Y(\tau)}$, the optimization over $\alpha$ of the right-hand side of Eq. (\ref{eq:info_imbalance_inequality}) takes the form of a \emph{minimum entropy} protocol, where we aim to maximize the predictability of the future distances in the space of the effect variable, when we consider points that were close in the past according to both the effect and the causal variables.

\section{Other approaches for causality detection}

\subsection{Measure \emph{L}} 

In \cite{chicharro2009reliable_SI}, Chicharro and Andrzejak introduced a rank-based measure to detect unidirectional couplings in non-synchronizing conditions, lying on the idea of state space reconstruction using time-delay embeddings (in the following, $\widetilde{x}(t)$ and $\widetilde{y}(t)$).
To write the expression of this statistical measure, we call $v_{tj}$ and $w_{tj}$ the time indices of the $j$-th nearest neighbors of $\widetilde{x}(t)$ and $\widetilde{y}(t)$, respectively, from which we exclude temporal neighbors within a window of size $W$ (requiring $\mid v_{tj}-t\mid > W$ and $\mid w_{tj}-t\mid > W$).
Distances are measured in the shadow manifolds according to Euclidean metrics, denoted by $d_X$ and $d_Y$.
Considering $k$ nearest neighbors for each point, the \emph{measure} $L$ is defined as
\begin{equation}
    L(X \mid Y)\doteq\frac{1}{N} \sum_{t=1}^N \frac{G_t(X)-G_t^k(X \mid Y)}{G_t(X)-G_t^k(X)}\,,
\end{equation}
where $G_t(X)$, $G_t^k(X)$ and $G_t^k(X\mid Y)$ are respectively the mean rank, the minimal mean rank and the $Y$-conditioned mean rank:  $G_t(X)=\frac{N}{2}$, $G_t^k(X)=\frac{k+1}{2}$ and $G_t^k(X\mid Y) = \frac{1}{k}\sum_{j=1}^k r^X_{t,w_{tj}}$ ($r_{tt'}^X$ denotes the distance rank of $\widetilde{x}(t')$ with respect to $\widetilde{x}(t)$). 
Taking only the first nearest neighbor ($k=1$), it is easy to show that the measure $L$ is related to the standard Information Imbalance by a straightforward linear relation:
\begin{subequations}
\begin{align}
    L(X \mid Y) &= \frac{1}{N} \sum_{t=1}^N \frac{N/2-G_t^1(X \mid Y)}{N/2-1} \\
    &= \frac{N}{N-2} - \frac{2}{N-2}\frac{1}{N}\sum_{t=1}^N G_t^1(X\mid Y) \\
    &= \frac{N}{N-2} - \frac{2}{N-2} \langle r_X \mid r_Y = 1 \rangle \\
    &= \frac{N}{N-2} \left[1 - \Delta(d_Y \rightarrow d_X) \right]\,. \label{eq:II_L_measure}
\end{align}
\end{subequations}
Using a generic number of neighbors $k$, this relation becomes
\begin{equation}
    L(X \mid Y) = \frac{N}{N-k-1}\left[1 -\Delta(d_Y\rightarrow d_X) \right]\,.
\end{equation}
We notice that the Information Imbalance of Eq. (\ref{eq:II_L_measure}) is referred to distance spaces with no relative time lag ($\tau = 0$), which is a major difference between state space reconstruction methods such as the measure $L$ and methods based on predictability.
In \cite{chicharro2009reliable_SI} the authors proposed to assess the presence of the coupling $X\rightarrow Y$ by checking whether the condition $\Delta L=L(X\mid Y)-L(Y\mid X)>0$ is verified.
In terms of the Information Imbalance measure, this translates into the inequality $\Delta(d_Y \rightarrow d_X) < \Delta(d_X \rightarrow d_Y)$.

For the analysis reported in Fig. 3 in the main text we chose the optimal values of the embedding time $\tau_e$ following the mutual information criterion, except for the case of the Lorenz 96 systems, where we found $\tau_e = 1$ to minimize the difference between the signals in the $X\rightarrow Y$ and $Y\rightarrow X$ couplings. 
We set $k = 5$ both for the R\"{o}ssler and the Lorenz systems, and $k = 20$ for the Lorenz 96 systems.

\subsection{Extended Granger causality} 

Given two one-dimensional signals $x(t)$ and $y(t)$, the Extended Granger Causality method \cite{Chen2004analyzing_SI} tries to assess wheteher $x$ has a causal influence on $y$ by studying which one among the two autoregressive models
\begin{subequations}\label{eq:EGC}
\begin{align}
    y(t+\tau_e) &= \sum_{j=1}^E\,\alpha_j\,y(t+(j-1)\tau_e) + \varepsilon_y\,, \\
    y(t+\tau_e) &= \sum_{j=1}^E\bigg[\beta_j\,y(t+(j-1)\tau_e) + \gamma_j\,x(t+(j-1)\tau_e)\bigg] + \varepsilon_{y\mid x}\,, \label{eq:EGC_2}
\end{align}
\end{subequations}
gives the best description of the unknown dynamics, performing several local regressions according to Eqs. (\ref{eq:EGC}) instead of a single global fit as in standard Granger causality.
Following ref. \cite{krakovska2018comparison_SI}, in all our analysis we carried out 200 local regressions.
Each local regression is carried out by sampling a random point in the space of time-delay embeddings and considering the neighbor points within a radius $\delta$ from the central one, as measured by the Euclidean distance in the space of the joint embeddings $(x(t), ...,\, x(t-(E-1)\tau_e), \,y(t), ...,\, y(t-(E-1)\tau_e)$. 
Equivalently, in this work we defined the neighborhood size by fixing the number of neighbors $k$.
The residuals estimated in each local regression are used to compute the Extended Granger Causality index $\Delta_{x\rightarrow y} = \langle 1 - \varepsilon_{y\mid x} / \varepsilon_x\rangle$, which identifies the presence of a coupling when statistically different from zero.

In the tests reported in Fig. 3 in the main text we 
chose $k$ using a single experiment at fixed $\varepsilon$ for each pair of systems, and then we used it to compute the extended Granger causality index for all the experiments.
We repeated these steps for several choices of the time-delay embedding parameters, selecting 
$k = 200$, $E = 3$ and $\tau_e = 1$ both for the R\"{o}ssler and the Lorenz systems, and $k = 500$, $E = 30$ and $\tau_e = 7$ for the Lorenz 96 systems.

As the method was proposed for time series only, we adapted it to the case where $x$ is a static variable by replacing Eq. (\ref{eq:EGC_2}) with 
\begin{equation}
    y(t+\tau_e) = \sum_{j=1}^E\,\beta_j\,y(t+(j-1)\tau_e) + \gamma\,x + \varepsilon_{y\mid x} 
\end{equation}
in order to apply it when $x$ is the duration of the comparison stimulus in the EEG experiment.
\noindent Since the single EEG time series do not satisfy the requirement of stationarity, similarly to our approach we constructed a data set for each $t$ and each participant using the different trials as independent realizations.

While in our method $\tau$ and $\tau_e$ are distinct parameters, the Extended Granger Causality approach is constructed with $\tau = \tau_e$.
For this reason, instead of studying the behavior of the Extended Granger Causality index as a function of the separation $\tau$ between the predictive window and the predicted point, we fixed $\tau_e$ and scanned the time $t$ of the predicted point, moved concurrently with the predictive window.
In the analysis carried out with our approach, by contrast, the predictive window was kept fixed after the stimulus onset / offset. 

In the EEG analysis we found the best results for $E = 3$ and setting $k$ to the total number of points available for each subject, namely performing a single global regression (insets of Fig. 4\emph{A} and \emph{B} in the main text).
Notably, with these parameters the Extended Granger Causality is equivalent to the standard Granger causality test.

\subsection{Convergent Cross Mapping} 

The method was implemented as described in ref. \cite{sugihara2012detecting_SI}, using neighborhoods of size $E + 1$ in the shadow manifolds of $x_1$ and $y_1$ to compute the cross mapping coefficients. 
The cross-map skill, quantified by the Pearson correlation $\rho$ between the reconstructed and the ground-truth points in spaces $x_1$ and $y_1$, was studied as a function of the length $L$ of the trajectory in order to select $\rho$ after convergence. 
In particular, we observed convergence at $L = 30000$ for the identical R\"{o}ssler systems ($E = 3,\, \tau_e  = 5$), $L = 20000$ for the different R\"{o}ssler systems ($E = 3,\, \tau_e  = 5$), $L = 40000$ for the bidirectionally coupled Lorenz systems ($E = 3,\, \tau_e  = 3$) and $L = 50000$ for the Lorenz 96 systems ($E = 30,\, \tau_e  = 7$).
The values of $\tau_e$ were chosen as the first minima of the lagged mutual information \cite{fraser1986independent_SI}.

\subsection{Transfer Entropy}

We employed the formulation provided by Palu\v{s} and Vejmelka in ref. \cite{palus2007directionality_SI}, which presents the measure in the $X\rightarrow Y$ direction through the conditional mutual information $I\big(\tilde{x}(0);\,\tilde{y}(\tau)\mid \tilde{y}(0)\big)$.
We implemented this measure with the conditional mutual information estimator proposed in ref. \cite{mesner2021conditional_SI} and implemented in the Python library knncmi, setting $k = 3$ for all the tested systems and using the following time-delay embedding parameters: $E=3$, $\tau_e=1$ for the R\"{o}ssler and Lorenz systems, and $E=30$, $\tau_e=1$ for the Lorenz 96 systems.
The time lag was set to $\tau = 20$ for the R\"{o}ssler systems, $\tau=5$ for the Lorenz systems and $\tau=30$ for the Lorenz 96 systems.

\section{Generalization of the Imbalance Gain to three systems}

The Imbalance Gain in direction $X\rightarrow Y$ defined in Eq. (3) of the main text can be generalized in order to take into account the effects induced by a third system $Z$:
\begin{equation}
    \delta \Delta ( X\rightarrow Y \mid Z) \doteq 1 - \frac{\underset{\alpha_X,\,\alpha_Z}{\min}\,\Delta\big(d_{\alpha_X X(0),\,\alpha_Z Z(0),\,Y(0)} \rightarrow d_{Y(\tau)} \big)}{\underset{\alpha_Z}{\min}\,\Delta\big(d_{\alpha_Z Z(0),\,Y(0)} \rightarrow d_{Y(\tau)}\big)}\,.
\end{equation}
We call this measure \emph{conditional} Imbalance Gain, as it quantifies how much adding the information of $X$ at time $t=0$ can improve the predictability of $Y$ at time $t=\tau$, given a baseline prediction which already includes the third system $Z$ at time $t=0$.
Similarly to the two-system case, we claim that $X$ causes $Y$ given $Z$ when $\delta \Delta ( X\rightarrow Y \mid Z) > 0$.
The scaling parameter $\alpha_X$ plays the same role of $\alpha$ in Eq. (3), while $\alpha_Z$ is an additional optimization parameter which scales the variables of system $Z$.
The optimal self-prediction in the unconditoned case, $\Delta(\alpha=0)$, is here replaced by $\min_{\alpha_Z}\,\Delta\big(d_{\alpha_Z Z(0),\,Y(0)} \rightarrow d_{Y(\tau)}\big)$. 
If the Imbalance Gain of Eq. (3) can be seen as a generalization of Granger causality, the conditional Imbalance Gain corresponds to its multivariate formulations (see for example refs.  \cite{Chen2004analyzing_SI,runge2018causal_SI}).

\begin{figure}
\centering
\includegraphics[width=15.8cm]{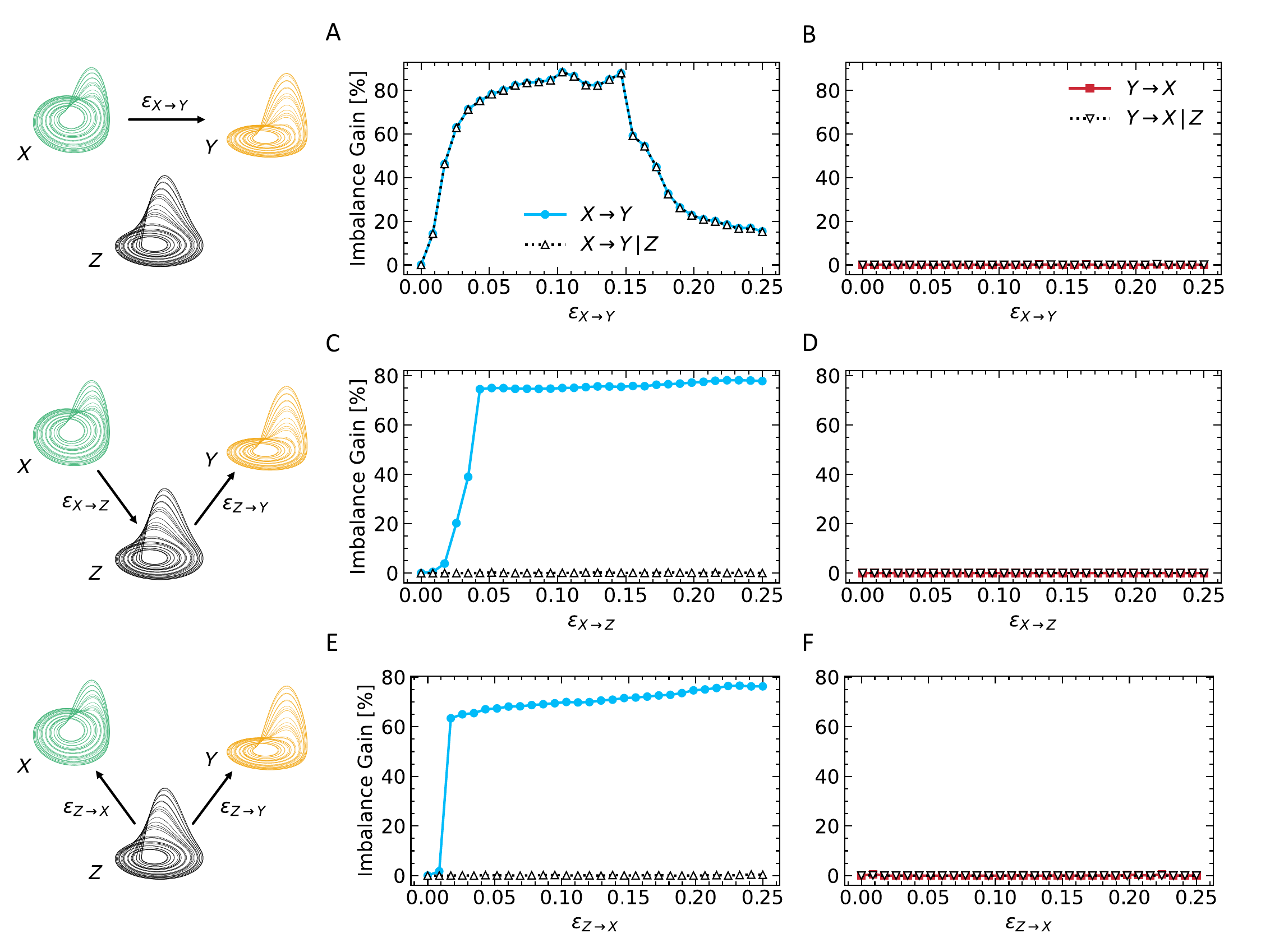}
\caption{Comparison between unconditioned and conditional Imbalance Gains, for three different R\"{o}ssler systems $X$, $Y$ and $Z$. \emph{(A),(B)}: Unidirectional coupling $X\rightarrow Y$, $Z$ uncoupled. 
\emph{(C),(D)}: Unidirectional couplings $X\rightarrow Z$ and $Z\rightarrow Y$, with $\varepsilon_{Z\rightarrow Y} = 0.05$.
\emph{(E),(F)}: Unidirectional couplings $Z\rightarrow X$ and $Z\rightarrow Y$, with $\varepsilon_{Z\rightarrow Y} = 0.05$. 
Each point was obtained by averaging 20 different independent estimates of the Imbalance Gain.
The minimization over $\alpha$ in the unconditioned measure was carried out using a grid of 50 evenly spaced points in $[0,0.25]$, while the minimization over $\alpha_X$ and $\alpha_Z$ in the conditional Imbalance Gain was performed using a two-dimensional grid with $50\times 50$ points in $[0,0.25] \times [0,0.25]$.
}
\label{fig:fig_cond}
\end{figure}

We tested the conditional measure on  three different Rössler systems, with parameters $\omega_X = 1.015$, $\omega_Y = 0.985$ and $\omega_Z = 1.005$, considering the following coupling scenarios (one for each row of Fig. \ref{fig:fig_cond}):

\begin{enumerate}
\item $X$ and $Y$ are unidirectionally coupled and $Z$ interacts neither with X nor with $Y$;
\item $X$ and $Y$ are indirectly coupled through $Z$;
\item $X$ and $Y$ are neither directly nor indirectly coupled and $Z$ is a common driver.
\end{enumerate}

\noindent The results reported in Fig. \ref{fig:fig_cond} were obtained using all the coordinates of the three systems and averaging both the unconditioned and the conditional measures over 20 independent estimates.
Similarly to the unconditioned case (see Materials and Methods - Average Imbalance Gain), in the conditional measure we averaged the $\alpha_X$-dependent Imbalance Gains obtained by optimizing only the $\alpha_Z$ parameter,
\begin{equation}
    \delta \Delta ( X\rightarrow Y \mid Z)(\alpha_X) \doteq 1 - \frac{\underset{\,\alpha_Z}{\min}\,\Delta\big(d_{\alpha_X X(0),\,\alpha_Z Z(0),\,Y(0)} \rightarrow d_{Y(\tau)} \big)}{\underset{\alpha_Z}{\min}\,\Delta\big(d_{\alpha_Z Z(0),\,Y(0)} \rightarrow d_{Y(\tau)}\big)}\,,
\end{equation}
inferring a single optimal $\alpha_X$ for all the independent estimates.

While in scenario 1) the conditional Imbalance Gain $\delta \Delta(X\rightarrow Y \mid Z)$ coincides with the standard measure $\delta \Delta(X\rightarrow Y)$  (panels A and B), in scenarios 2) and 3) it allows to clearly understand that X and Y are not directly coupled. 
This shows that our approach allows to exclude unambiguously the presence of a direct causal link from $X$ to $Y$ both when a common driver is present and when the link is indirect, opening up to applications in causal network reconstruction for time-dependent variables \cite{runge2018causal_SI}.

\section{Stability with respect to noise of the Imbalance Gain} 

\begin{figure}
\centering
\includegraphics[width=14.8cm]{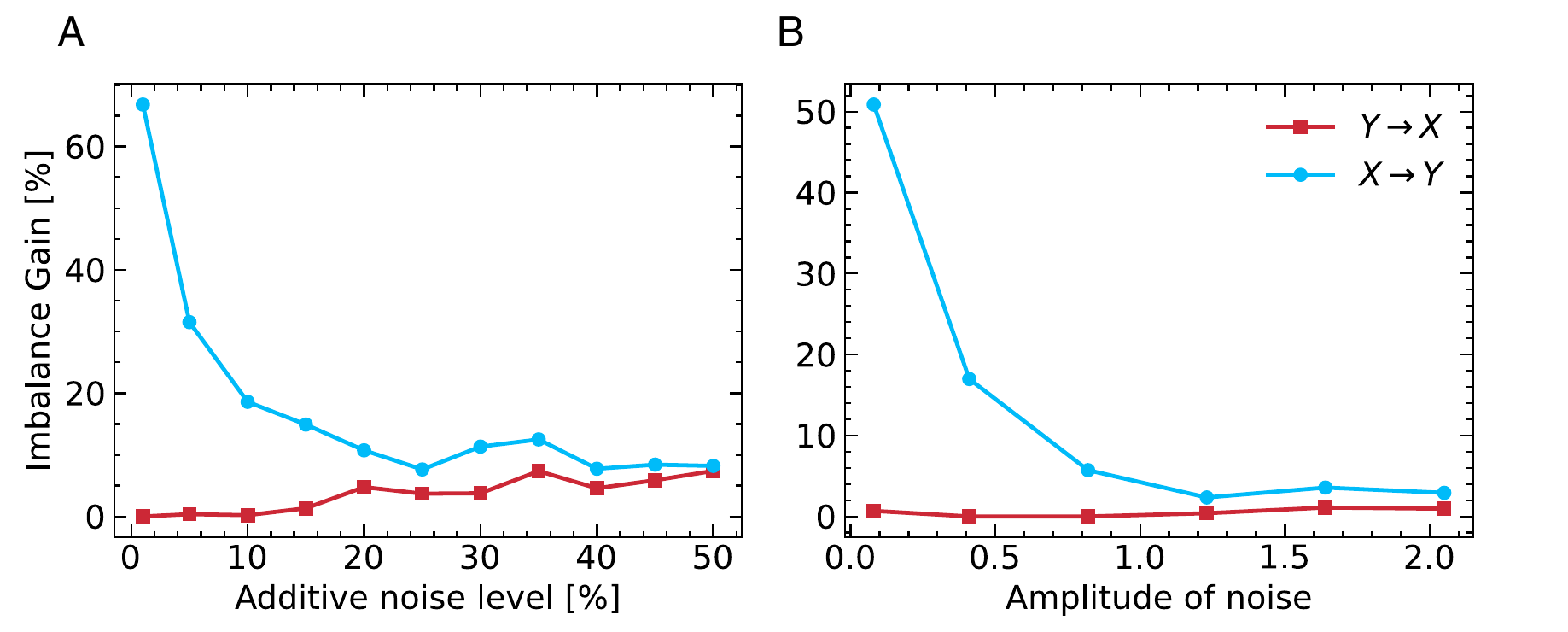}
\caption{Effect of noise on the Imbalance Gain, measured on the unidirectionally coupled R\"{o}ssler systems ($X\rightarrow Y$) with coupling $\varepsilon = 0.1035$. (\emph{A}) White noise was added a posteriori and independently to each dynamical variable, choosing its amplitude as a fraction (along the $x$-axis) of the standard deviation of the entire trajectory in the corresponding direction. (\emph{B}) Independent white noises were added during the integration of the dynamical systems, only to the variables $x_1$ and $y_1$, with amplitude reported along the $x$-axis. Both the tests were carried out with $k=1$ and $\tau = 5$.}
\label{fig:fig_noise}
\end{figure}

Real-world time series can be ``noisy'' as a result of measurement errors or as a consequence of a stochastic underlying dynamics.
We investigated these two scenarios separately, both adding noise \emph{a posteriori} to the deterministic trajectories (Fig. \ref{fig:fig_noise}\emph{A}) and by adding it at each integration step (Fig. \ref{fig:fig_noise}\emph{B}).
In the latter case we only considered the case of Gaussian and uncorrelated (white) noise, which turns the dynamical equations into Langevin-like equations. The use of correlated noise may hinder the performance of the method and is expected to affect the behavior of the Imbalance Gain as a function of $\tau$, depending on the characteristic time-scales of the noise.
To avoid mixing the effect of noise with other nuisance sources, we computed the Imbalance Gain considering all the coordinates of the two systems.

\begin{figure}
\centering
\includegraphics[width=14.8cm]{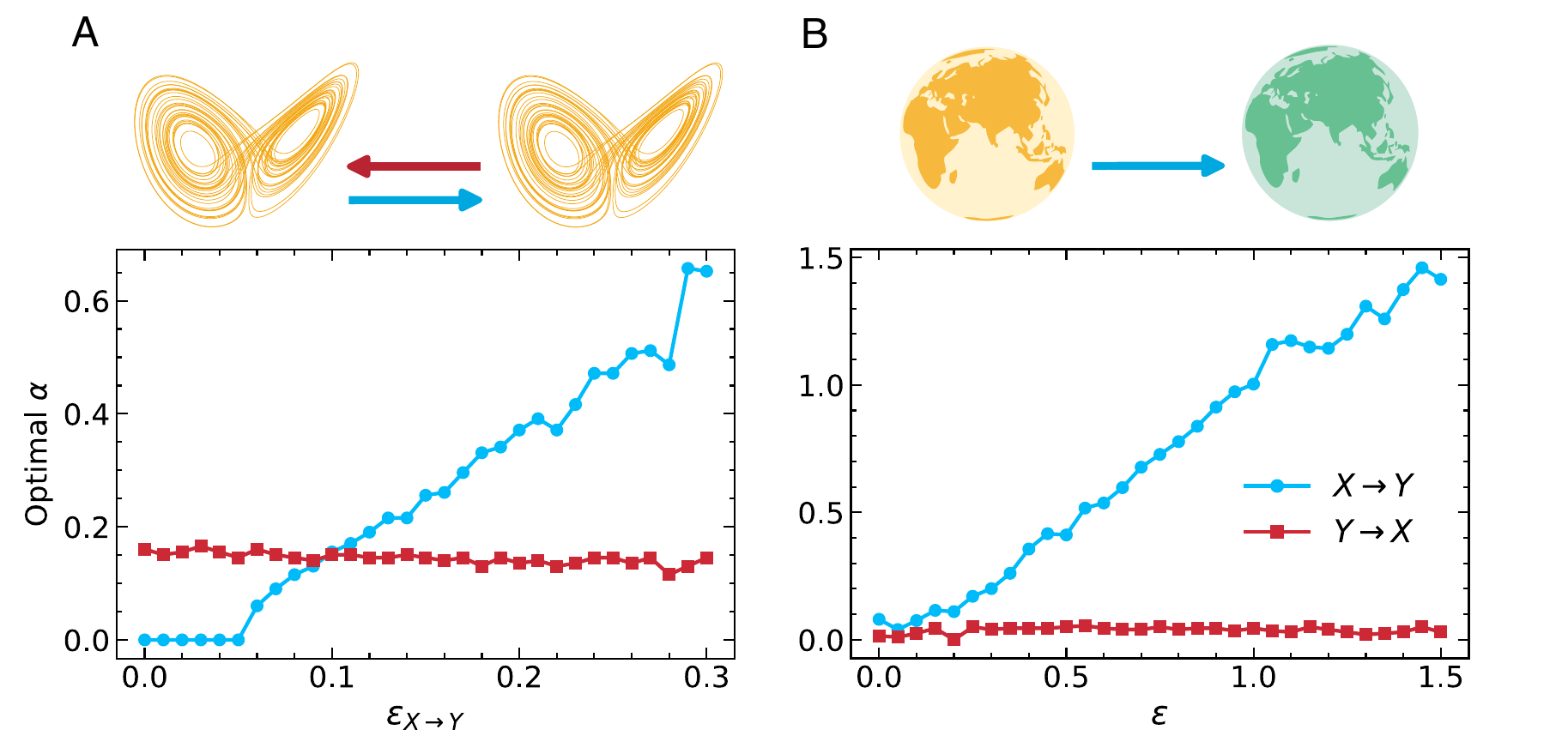}
\caption{Optimal $\alpha$ as a function of the coupling strength in direction $X \rightarrow Y$ for (\emph{A}) the bidirectionally coupled Lorenz systems and (\emph{B}) the unidirectionally coupled Lorenz 96 systems, using time-delay embeddings with the same parameters employed in the main text. The minimization was carried out over a grid of 300 values evenly spaced in range $\left[0,1.5\right]$.}
\label{fig:alpha_minimum_SI}
\end{figure}

As expected, in both scenarios the signal in the $X\rightarrow Y$ direction appears to weaken as the noise level increases. 
In direction $Y\rightarrow X$, where the coupling is absent, the signal is negligible up to a 15\% noise level in the first case (Fig. \ref{fig:fig_noise}\emph{A}), while the Imbalance Gain remains close to zero for all the tested noise amplitudes in the second scenario.
We interpret this difference as follows.
When the noise is added a posteriori on a deterministic signal, its effect is to reduce the information content of the distance space $d_{X(0)}$  with respect to $d_{X(\tau)}$, so that the former is not anymore the most informative metric that one can build with respect to the latter by using the dynamic variables at time $t =0$.
In this scenario the self-prediction of $X$ becomes non-optimal, leaving the possibility of improvements by including the information on $Y(0)$ in the distance space $d_{\alpha Y(0),\,X(0)}$. As Fig. \ref{fig:fig_noise}\emph{A} shows, the reduction of information induced by measurement noise in $d_{X(0)}$ results
in a nonzero Imbalance Gain in direction $Y\rightarrow X$.
When the underlying dynamics is stochastic, instead, the effect of noise is to reduce the predictability of $d_{X(\tau)}$ given the knowledge of $d_{X(0)}$, because the dynamics of $X$ is altered and the system exhibits a shortened memory.
In this context, the distance space $d_{X(0)}$ still remains the most informative that one can construct to make a prediction about $d_{X(\tau)}$ with the dynamic variables at time {$t=0$.
As a consequence, the information about $Y(0)$ keeps being redundant as in the case of a deterministic dynamics, and in Fig. \ref{fig:fig_noise}\emph{B} we observe consistently that the signal in direction $Y\rightarrow X$ is only slightly affected by the amplitude of the integration noise.

\section{Role of the optimization parameter}

In the presentation of the method we suggested that the scaling parameter $\alpha$ in the distance $d_{\alpha X(0),Y(0)}$ is necessary to account for the magnitude of the coupling strength in direction $X\rightarrow Y$. 
In Fig. \ref{fig:alpha_minimum_SI} we provide a numerical evidence of this claim by showing that for the dynamical systems not affected by synchronization (namely the bidirectionally coupled Lorenz systems and the unidirectionally coupled Lorenz 96 systems of Fig. 3) the optimal value of $\alpha$ in the minimum is a monotonically increasing function of the coupling strength, with small fluctuations likely due to statistical noise.
This property allows to use the optimal values of $\alpha$ for comparing different coupling strengths in two scenarios involving the same dynamical systems, but it does not imply that the absolute coupling strength can be inferred from the knowledge of the optimal $\alpha$.
As a simple counterexample, an arbitrary re-definition of the units of $X$ by a scaling factor 2 ($X\mapsto 2 X$) would scale the optimal $\alpha$ by a factor 1/2 ($\alpha\mapsto 1/2\,\alpha$), even if the coupling strength in direction $X\rightarrow Y$ is untouched.
The optimal scaling parameters displayed in Fig. \ref{fig:alpha_minimum_SI} were obtained by using 20 independent realizations of the same dynamics, as described in the Materials and Methods - Average Imbalance Gain.

\section{Assessing the statistical significance of the Imbalance Gain for a single estimate}

Due to finite-sample effects, the Imbalance Gain (Eq. (3) in the main text) can be small but nonzero even in absence of causality.
This requires a quantitative approach to evaluate whether a single estimate of the Imbalance Gain is compatible with zero.

The statistical significance of our measure can be assessed by a permutation test which generates a null distribution under the null hypothesis of ``absence of causality''.
If we consider the direction $X\rightarrow Y$, we can generate the null distribution of the Imbalance Gain $\delta \Delta (X\rightarrow Y)$ under the hypothesis $H_0$ ``$X$ does not cause $Y$'' by applying random permutations to the data point indices (i.e. the trajectories in the ensemble) in the only space $X$.
For example, if trajectory $i$ includes the components $X^i = (x^i_1,\,x^i_2,\,x^i_3)$ and $Y^i = (y^i_1,\,y^i_2,\,y^i_3)$, we denote by $\sigma$ a permutation that only affects the trajectory index in the $X$ space: $X^{\sigma(i)} = (x^{\sigma(i)}_1,\,x^{\sigma(i)}_2,\,x^{\sigma(i)}_3)$. The Imbalance Gain in direction $X\rightarrow Y$ under such a random permutation,
\begin{equation}\label{eq:permutations}
    \delta \Delta^\sigma (X\rightarrow Y)\doteq \frac{\Delta\big(d_{Y(0)}\rightarrow d_{Y(\tau)}\big) - \min_\alpha\Delta\big(d_{\alpha X(0)^\sigma,Y(0)}\rightarrow d_{Y(\tau)}\big)}{\Delta\big(d_{Y(0)}\rightarrow d_{Y(\tau)}\big)}\,,
\end{equation}
is by construction compatible with zero independently of the ground-truth coupling scenario.
Indeed, permuting the indices results in a randomization of the distances in space $d_{X(0)}$, eliminating any information that this space could provide about distances measured in $d_{Y(\tau)}$.
Therefore, computing Eq. (\ref{eq:permutations}) under many random permutations in space $X(0)$ allows sampling the null distribution $p_{H_0}\left(\delta \Delta\right)$, from which we can compute the $p\,$-value of the actual Imbalance Gain and accept (reject) the null hypothesis when $p$ is larger (smaller) than a chosen confidence.
The same approach can be applied in direction $Y\rightarrow X$ by permuting the point indices within the $Y$ space only.

We measured the statistical significance of single Imbalance Gain estimates for the unidirectionally coupled Lorenz 96 systems (the same employed in the comparisons of Fig. 3 in the main text). 
As in the main text, we considered 31 values of the coupling strength $\varepsilon$ equally spaced in the interval $[0,1.5]$.
Tables \ref{tab:confusion_matrix}\emph{A} and \emph{B} report the confusion matrices associated to the outcome of the statistical test, both using all the 40 coordinates of the systems and with time-delay embeddings ($E=30$).
The rate of false positive detections is below $10\%$ in both cases. Using time-delay embeddings the rate of false negative detections increases to 16.7\%, but all the wrong detections are limited to the weak coupling regime ($\varepsilon < 0.25$).

\section{Assessing the false positive rate of repeated and independent estimates}

For each measure used in the comparison of Fig. 3, the false positive rate was assessed over systems coupled in direction $X\rightarrow Y$ by testing the direction without causal link ($Y\rightarrow X$) according to the following steps:
\begin{enumerate}
\item The measure in direction $Y\rightarrow X$ was repeatedly computed over 20 independent realizations of the dynamics, thus providing 20 independent estimates $\{m_i \}$ ($i=1, ... ,20$). The independent realizations were obtained by integrating the dynamical equations from different initial conditions and discarding the first 100000 points of each integration.
\item For each value of $\varepsilon$, the quantity $t = \langle m\rangle /\, \text{SE}\left[\langle m\rangle \right]$ was computed, where $\text{SE}\left[ \langle m\rangle \right] = \text{Std} \left[ m \right] /\, \sqrt{20}$ denotes the standard error of the mean. 
This statistic asymptotically follows a Student’s $t$-distribution with $20 - 1$ degrees of freedom.
\item In order to assess if $\langle m\rangle$ is compatible with zero, a Student’s $t$-test employing the estimate of $t$ in the previous point was carried out. 
According to this test, the null hypothesis “the population mean of the causal measure $m$ is zero in direction $Y\rightarrow X$” is rejected when $t > $ threshold. 
We selected the threshold value in order to reject the null hypothesis with $p$-value $p<0.001$, but we also studied the sensitivity of the test as a function of $p$, as reported in Fig. \ref{fig:fig_FPR}. 
Since $\langle m\rangle$ is non-negative for all the causality measures, we employed a one-tailed test where $p$ corresponds to the probability of sampling a value larger than $t$, namely we only considered the right tail of the distribution.
\end{enumerate}
After performing the $t$-test for each value of $\varepsilon$, the false positive rate was computed as the fraction of $\varepsilon$ values resulting in a rejection of the null hypothesis.

\section{EEG data}

\subsection{Experimental details}

Volunteers underwent six experimental blocks, each composed by 60 trials. 
In each trial, a visual stimulus, varying in display time trial by trial, was presented at the center of the screen (Samsung 22\,'' SyncMaster 2233RZ, 1680x1050 pixels resolution, 120 Hz refresh rate, viewing distance 60 cm, grey background color). 
This stimulus, called comparison, was a sinusoidal annulus grating (5 degrees of visual angle outer-circle diameter, 1 degree of visual angle inner-circle diameter, 1 cycle per degree, 100\% contrast) composed by 5 concentric parts of equal area slowly and rigidly rotating around the center of the screen ($\sim0.6$ rad/s with alternating rotation senses, either clockwise or counterclockwise). 
The comparison stimulus could be displayed for 0.3, 0.4, 0.5, 0.6, 0.7, 0.8 or 0.9 seconds in each trial and it was followed (after an interval uniformly distributed between 0.5 and 1 second) by an auditory stimulus of 0.6 seconds (burst of white noise delivered through Etymotic Research ER1 Insert Earphones). 
Each stimulus duration was tested 10 times in each block.
Volunteers were instructed to maintain their gaze on a fixation cross (0.2 degrees of visual angle in diameter) presented at the center of the screen throughout the experiment and to report, by pressing a button on the keyboard, whether the first or the second stimulus was presented for longer time. 
After providing the response the next trial started automatically after an interval uniformly distributed between 0.5 and 1 second. 
Volunteers received no feedback on their performance.
Volunteers performed the task comfortably sitting in front of the apparatus while we recorded their EEG activity using 64 active electrodes (arranged in a 10-20 layout, GND in Fpz and REF in FCz, cap model: actiCAP slim) BrainProducts actiCHamp Plus (Brain Products GmbH, Gilching, Germany) with 2048 Hz sampling frequency. No filtering was applied during the recording}. 
In order to minimize head motion and blink artifacts, volunteers placed their head on a chin rest and they were instructed to blink when providing the response. 

\subsection{Data preprocessing}

EEG data preprocessing was performed using EEGLAB toolbox \cite{DELORME20049_SI} in Matlab (Mathworks Inc.). 
First, we removed motion and blink artifacts using automatic labeling of independent components (IC) using the ICLabel toolbox \cite{PIONTONACHINI2019181_SI}. 
Only ICs with a probability higher then 90\% of being correctly classified as muscle or eye artifact were removed. 
Then, the continuous EEG signal was downsampled to 250 Hz sampling rate, referenced to a common average reference and filtered using a 4\textsuperscript{th}-order Butterworth band-pass filter with range 0.1-40 Hz. 
The EEG data corresponding to the comparison stimulus was epoched starting from 200 ms before its onset to 500 ms after its offset. 
These epochs were detrended, and were examined to remove any remaining artifacts. 
We automatically removed epochs based on signal amplitude threshold, removing all trials in which the EEG amplitude exceeded $\pm50\,\mu$V. 
All the remaining trials were visually inspected to detect and remove the ones containing any residual artifacts. 
A total of $\sim 0.1 \%$ of trials were discarded. 
Additionally, we excluded from further analyses temporal channels (T7, T8, TP7, TP8, TP9, TP10, P7, P8, FT7, FT8, FT9, FT10).
We constructed then two data sets corresponding to the onset and offset period of comparison presentation, available at the link reported in the main text (Materials and Methods section - Data, Materials and software availability).
The onset data set contained EEG traces of 300 ms length starting from stimulus onset baseline corrected using 44 ms pre-stimulus period.
The offset data set consisted of EEG traces of 500 ms length starting from stimulus offset baseline corrected using the 88 ms window centered at stimulus offset (44 ms pre-offset and 44 ms post-offset period).

\subsection{Validation of the Imbalance Gain measure with a dichotomous causal variable}

As an additional test of our measure, we carried out a causal analysis where $X$ is a dichotomous variable discriminating between trials where the stimulus is presented after a fixed time interval ($\Delta t = 80$ ms) and trials where the visual stimulus is not presented.
Since the experiment did not include control trials where the visual stimulus was not presented to participants, the data set for this analysis was built from the original EEG time series with the following steps:
\begin{itemize}
    \item for each participant half of the trials (labelled by $X=1$) where shifted by 120 ms, setting the initial time $t=0$ to -80 ms from the onset of the stimulus, while for the remaining trials (labelled by $X=0$) the initial time was set to -200 ms from stimulus onset;
    \item a baseline correction was carried out using the initial 44 ms of the new series, namely using the EEG signal in range $\left[ -200, -256 \right]$ ms from stimulus onset for the trials of class $X=0$, and the range $\left[ -80, -36 \right]$ ms for the series labelled by $X=0$.
\end{itemize}
By employing time-delay embeddings of 44 ms, a trivial causality between $X$ and the EEG signal should arise only for $\tau > 80 - 44$ ms, namely when the predicted windows cross the time point at which the trials of class $X=0$ and $X=1$ start to differentiate. Consistently, the Imbalance Gain shown in Fig. \ref{fig:fig_trivial_causality} becomes nonzero after the aforementioned time lag. The analysis was restricted to one of the occipital electrodes (POz), where the first cortical signature of a visual input should be detected \cite{dirusso2002cortical_SI}.

\section{Coupled dynamical systems with different time scales}

An interesting dynamical scenario occurs when the driver system is particularly faster or slower than the driven system. 
In order to study how the Imbalance Gain and the other causality measures behave in such a scenario, we extended the equations of the unidirectionally coupled Lorenz 96 systems with an additional parameter $\widetilde{\omega}$, which tunes the velocity of X:
\begin{subequations}
\begin{align}
    \dot{X} &= \widetilde{\omega} f(X) \,, \\
    \dot{Y} &=  g(Y) +  G(X,Y) \,.
\end{align}
\end{subequations}
The explicit expressions of the functions $f$, $g$ and $G$ are reported in the Materials and Methods - 
Details on the dynamical systems. 
The systems tested in the last column of Fig. 3 correspond to the specific case $\widetilde{\omega} = 1$. 
Increasing $\widetilde{\omega}$ is equivalent to increase the velocity of $X$ while keeping the velocity of $Y$ untouched.

In Fig. \ref{fig:fig_omega} we report the behaviors of the different causality measures as a function of the parameter $\widetilde{\omega}$.
For all the methods we employed the same parameters used for the Lorenz 96 systems of Fig. 3, and we computed errors over 20 independent estimates.
The critical observation, which makes the interpretation of the results nontrivial, is that the variation of $\widetilde{\omega}$ appears to drastically affect the properties of system $Y$, changing its Lyapunov exponents and its chaoticity.
Therefore, the effects of varying the relative velocity of $X$ and $Y$ on the causality measures are highly non-linear, as small changes of $\widetilde{\omega}$ can result in driven systems $Y$ with noticeably different dynamical properties.

Remarkably, also in this test we observe that all the causality measures, except the Imbalance Gain, are not able to recognize that causality  in the $Y\rightarrow X$ direction is absent, bringing to false-positive rates of almost 100\% (the false positives for the Imbalance Gain are at most 15\% if one uses time-delay embeddings, panel \emph{B}).


\clearpage

\clearpage

\clearpage

\clearpage
\begin{figure}
\centering
\includegraphics[width=15.8cm]{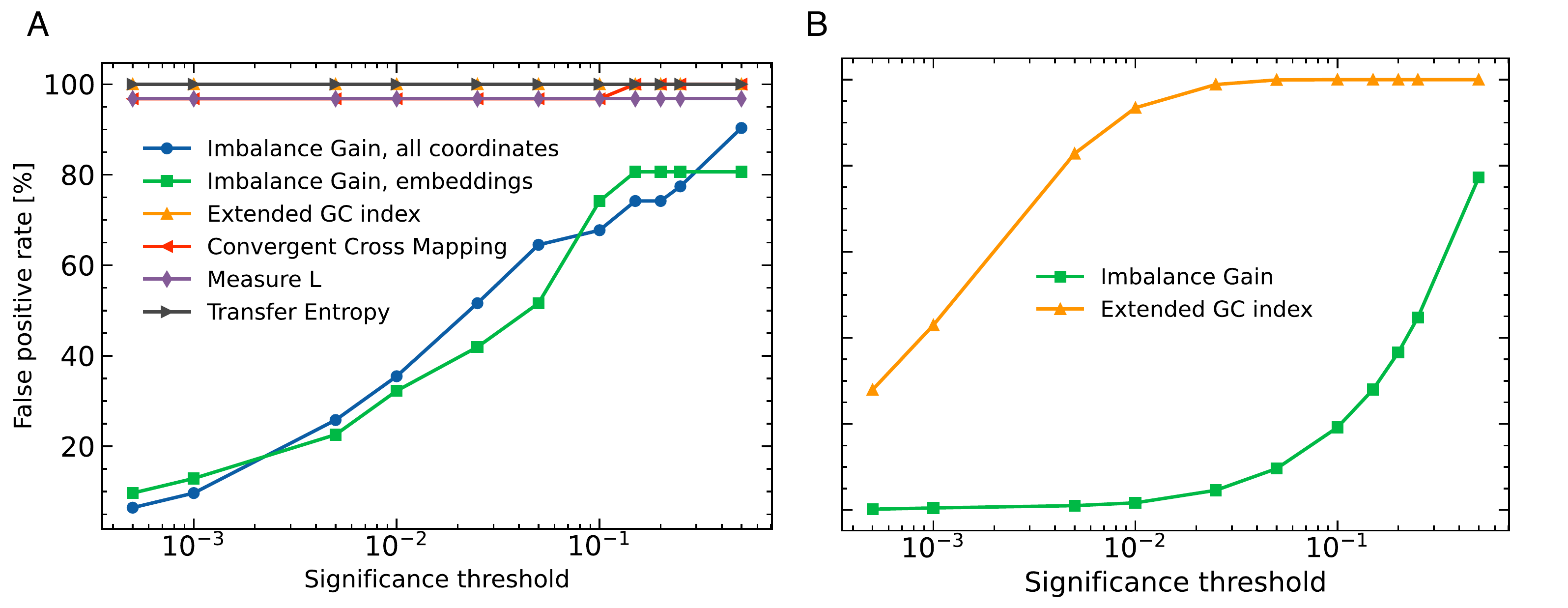}
\caption{False positive rates as a function of the significance threshold $\tilde{p}$, in two different scenarios where causality in the ground-truth dynamics is absent: (\emph{A}) in the tests $Y\rightarrow X$ for the coupled Lorenz 96 systems, and (\emph{B}) in the tests after stimulus onset in the EEG experiment, referred to the link between the duration of the stimulus and the EEG activity of the 51 channels. The null hypothesis of absence of causality is rejected for $p < \tilde{p}$.}
\label{fig:fig_FPR}
\end{figure}

\newpage
\begin{figure}
\centering
\includegraphics[width=10.1cm]{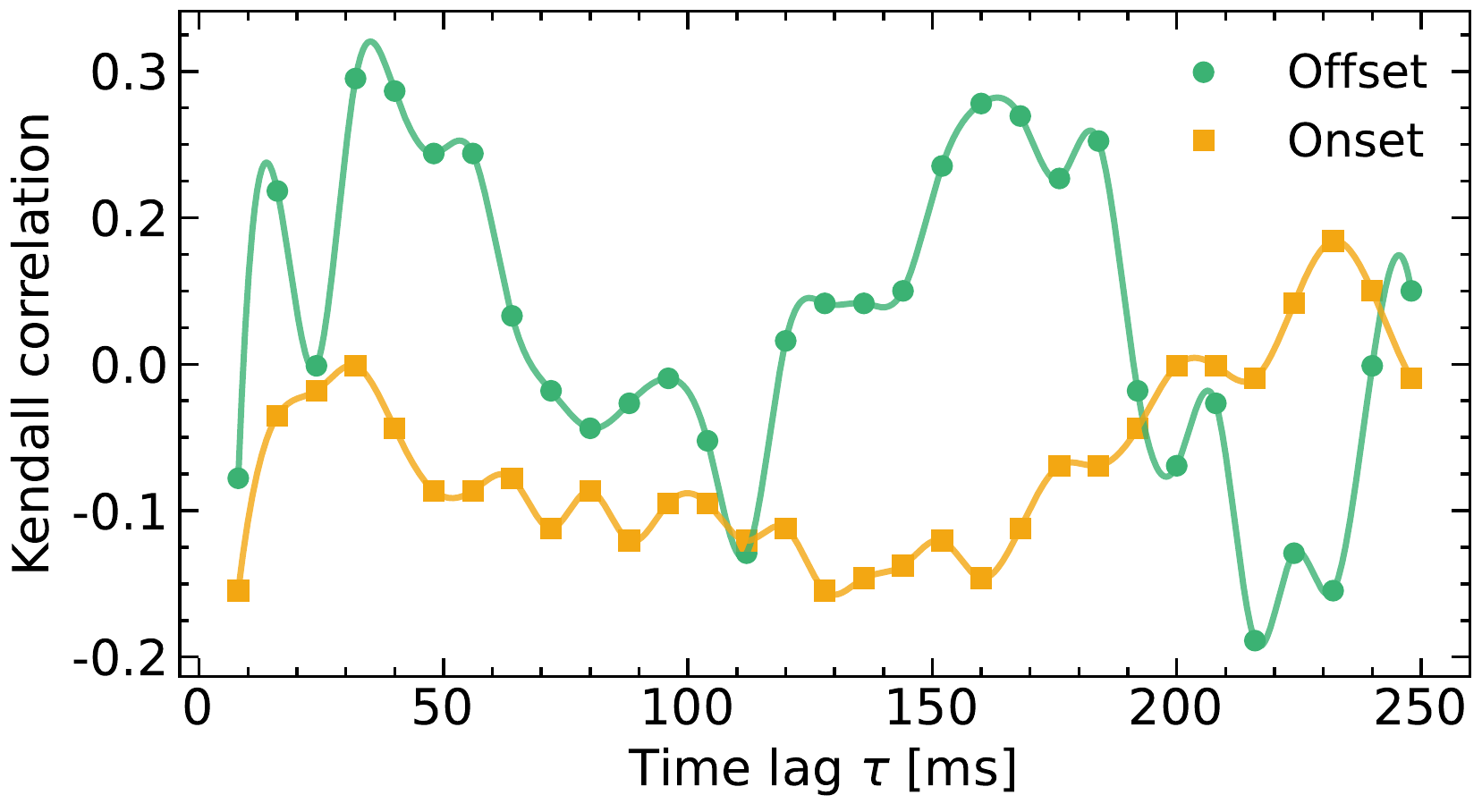}
\caption{Kendall correlation coefficient between the Imbalance Gains of the 19 participants in direction POz $\rightarrow$ Fz (see Fig. 4 in the main text) and their accuracies in the evaluation of the comparison stimulus' duration, as a function of the time lag $\tau$ from stimulus onset / offset.
Data were interpolated with a cubic spline.
The accuracy of each participant is defined as the number of correct answers divided by the total number of trials.}
\label{fig:fig_correlation}
\end{figure}

\clearpage
\begin{figure}
\centering
\includegraphics[width=10.8cm]{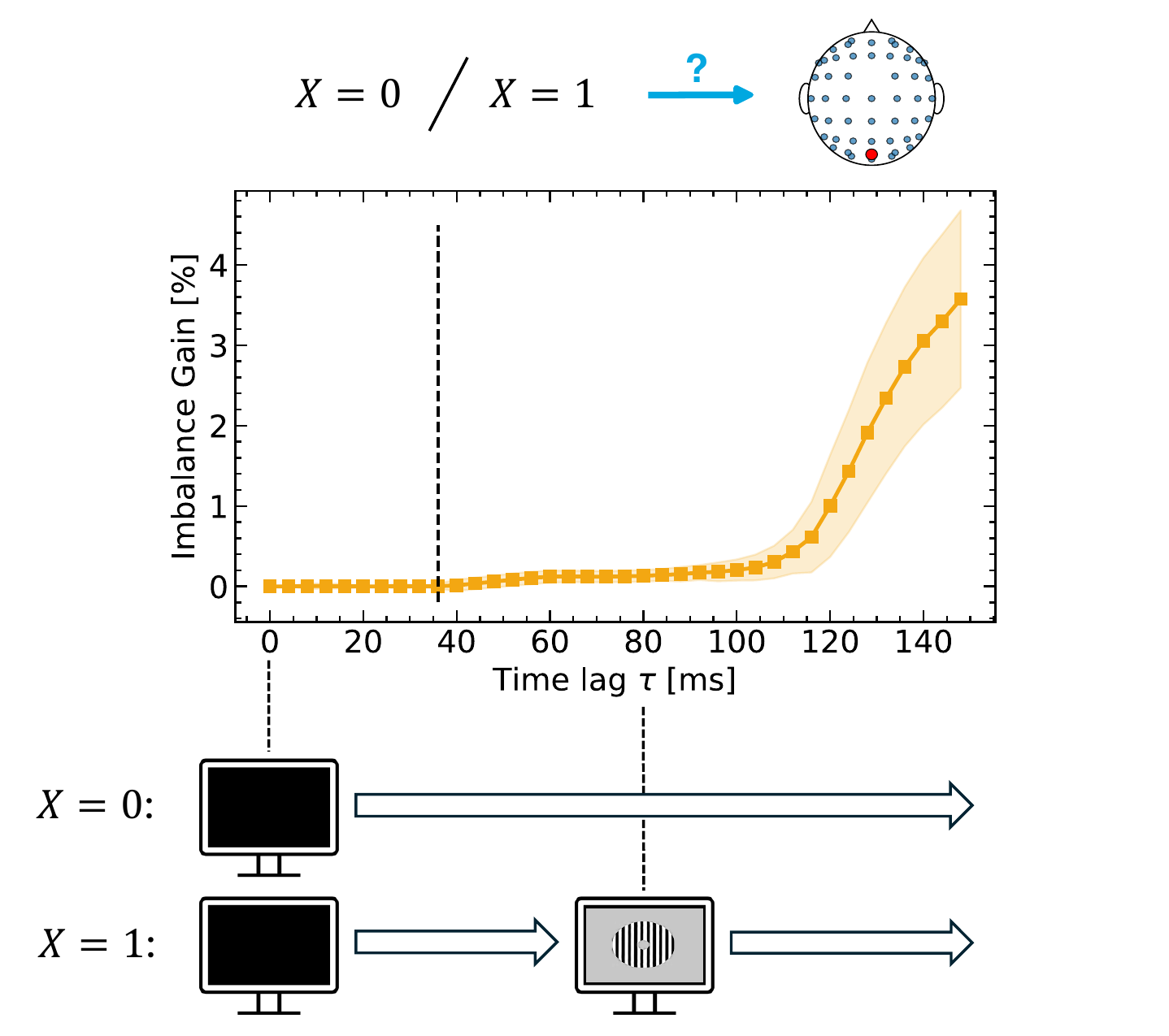}
\caption{Imbalance Gain as a function of the time lag $\tau$ to assess the presence of a trivial causal link between the dichotomous variable $X$ and the EEG activity of the occipital channel POz, using time windows of 44 ms. The variable $X$ discriminates between trials where the visual stimulus is presented after 80 ms and trials where the visual stimulus is not presented for any tested value of $\tau$. The dashed black line corresponds to the time lag $\tau = 80 - 44$ ms, from which the predicted EEG window includes the stimulus onset for $X=1$.}
\label{fig:fig_trivial_causality}
\end{figure}

\newpage
\begin{figure}
\centering
\includegraphics[width=17.5cm]{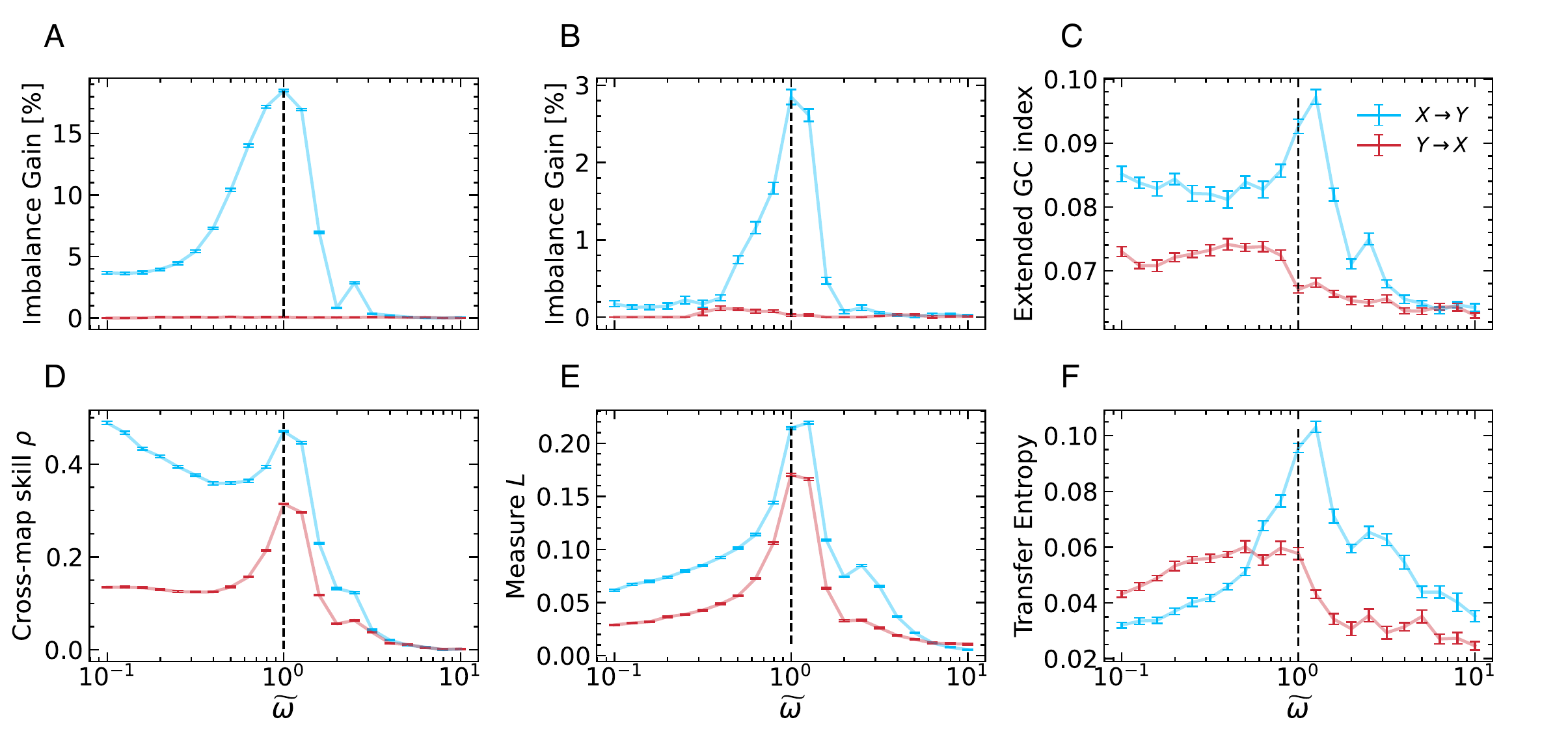}
\caption{Behaviors of the different causality measures against variations of the relative velocity between systems $X$ (driver) and $Y$ (driven), for two unidirectionally coupled Lorenz 96 systems. (\emph{A}): Imbalance Gain using all the coordinates, (\emph{B}): Imbalance Gain with time-delay embeddings, (\emph{C}): Extended Granger Causality, (\emph{D}): Convergent Cross Mapping, (\emph{E}): Measure L, (\emph{F}): Transfer Entropy.
}
\label{fig:fig_omega}
\end{figure}

\newpage
\begin{figure}
\centering
\includegraphics[width=12.5cm]{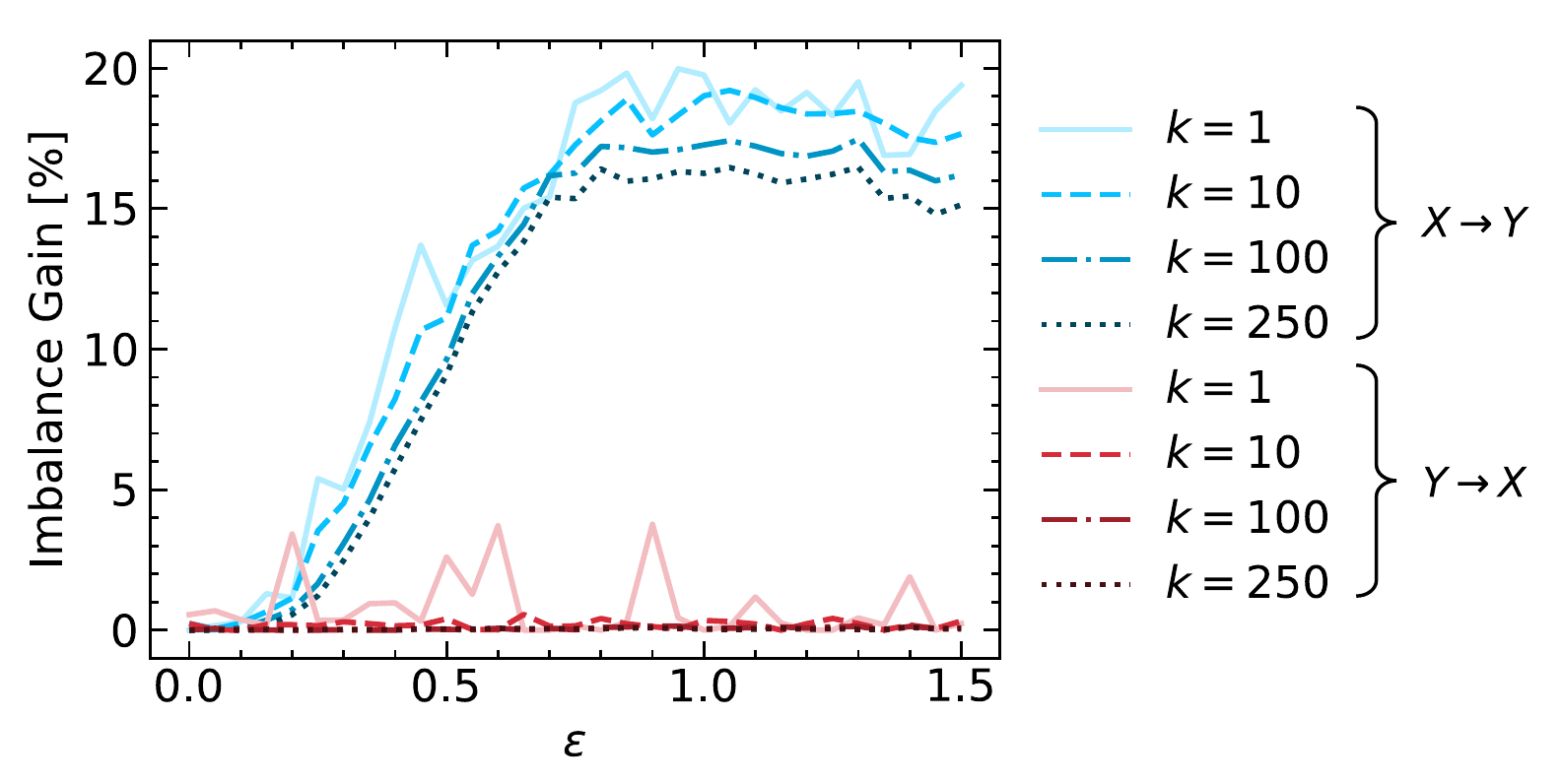}
\caption{Imbalance Gain as a function of the coupling strength $\varepsilon$ for the unidirectionally coupled Lorenz 96 systems ($F_X = 5$, $F_Y = 6$), using all the coordinates of $X$ and $Y$ and setting $\tau = 30$.
}
\end{figure}

\newpage
\begin{figure}
\centering
\includegraphics[width=12.5cm]{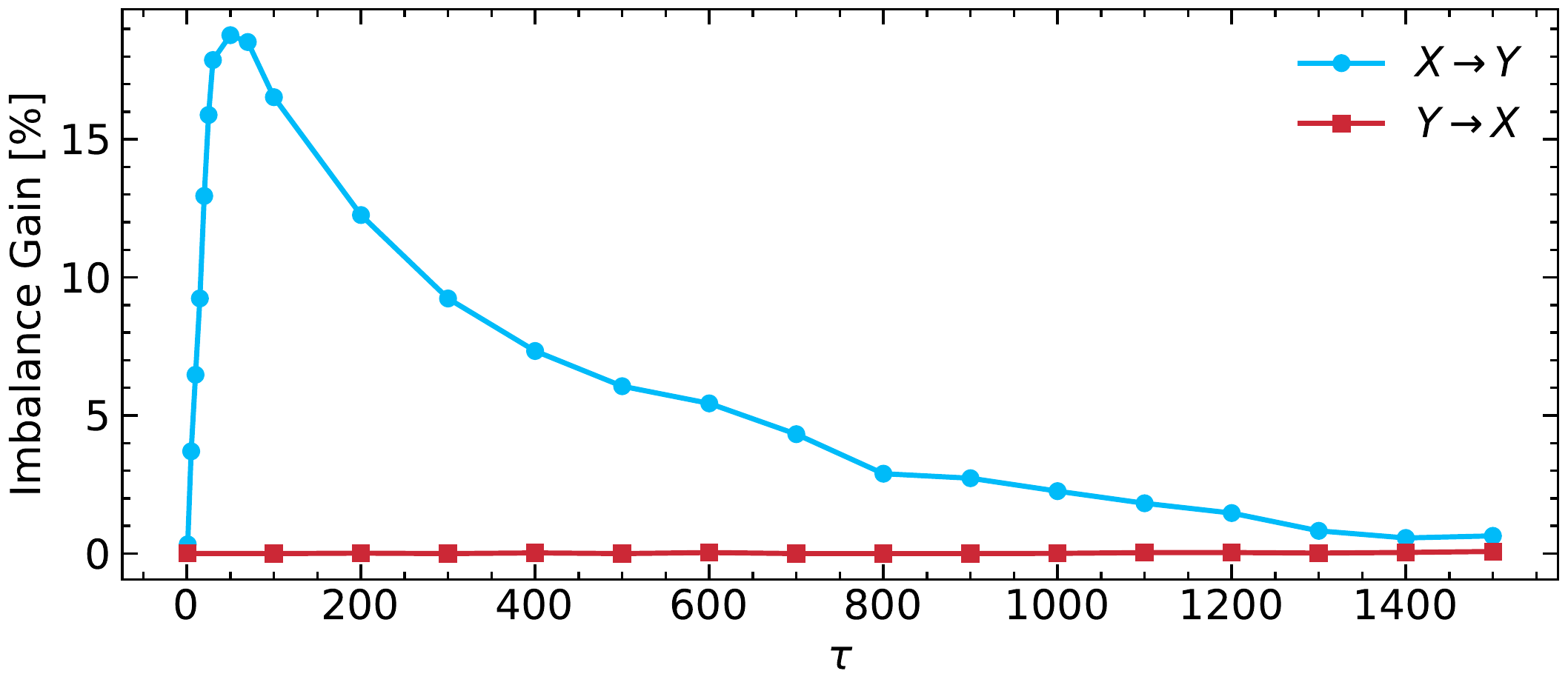}
\caption{Imbalance Gain as a function of the time lag $\tau$ for the unidirectionally coupled Lorenz 96 systems with $F_X = 5$, $F_Y = 6$ and $\varepsilon = 1$, using $k=20$.
}
\label{fig:fig_tau}
\end{figure}

\newpage
\begin{figure}
\centering
\includegraphics[width=12.5cm]{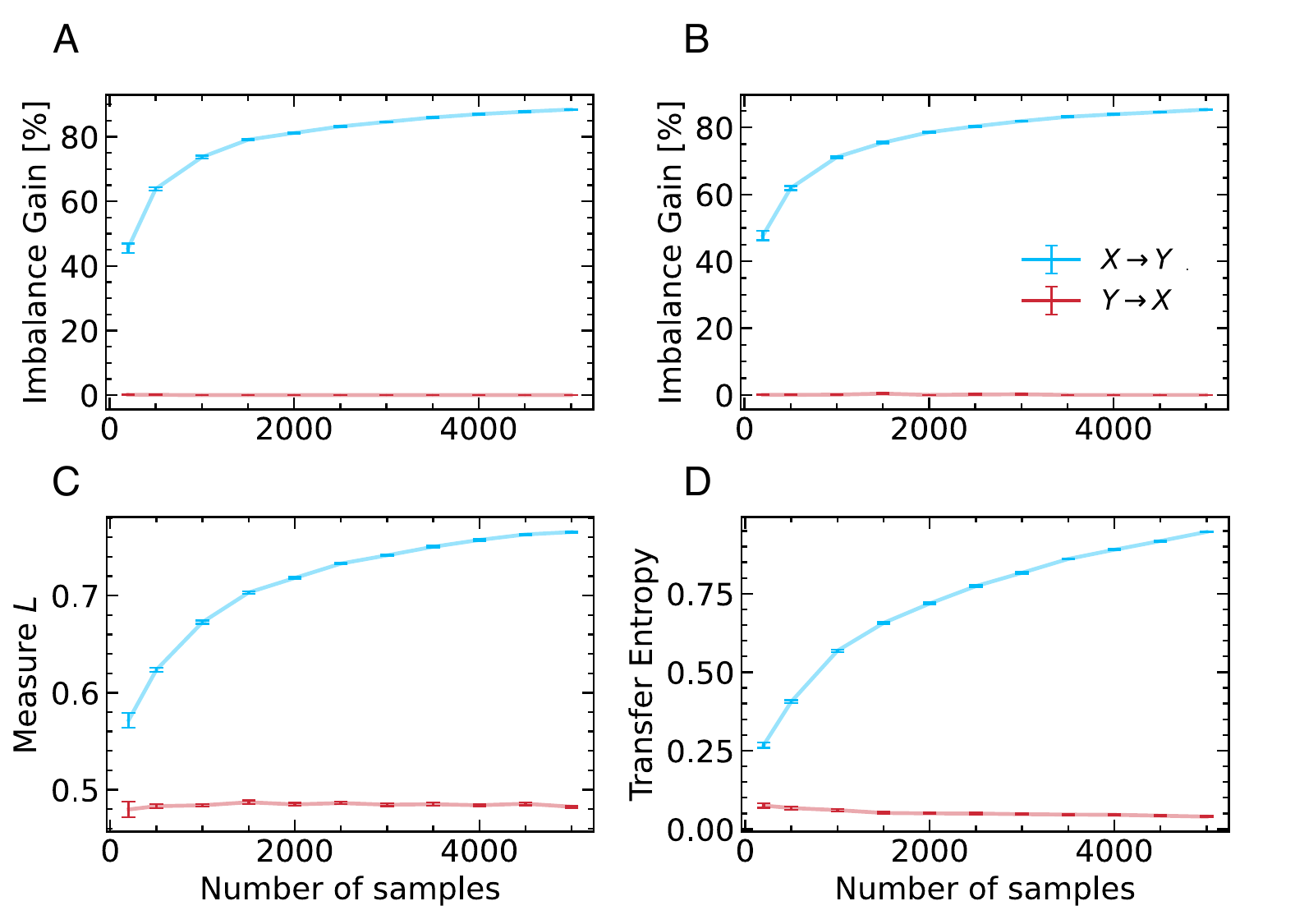}
\caption{Behaviors of different causality measures as a function of the number of samples $N$, for the different R\"{o}ssler systems coupled in direction $X\rightarrow Y$. (\emph{A}): Imbalance Gain using all the coordinates, (\emph{B}): Imbalance Gain with time-delay embeddings, (\emph{C}): Measure L, (\emph{D}): Transfer Entropy. 
The Convergent Cross Mapping approach was excluded from the comparison as it requires to monitor the convergence as a function of the number of samples, which may not be achieved when $N$ is small. Similarly, the Extended Granger Causality measure was not included in the comparison because the total number of points employed also depends on the number of points selected for the local regressions.
Each point and its standard error was computed over 20 independent estimates.
}
\label{fig:fig_N}
\end{figure}

\newpage
\begin{table}
    \begin{subtable}[h]{\textwidth}
        \centering
        \begin{tabular}{l r r}
         & Coupling (detected) & No coupling (detected) \\
        \hline
        Coupling (real) & 28/30 (93.3\%)& 2/30 (6.7\%)\\
        No coupling (real) & 1/32 (3.1\%) & 31/32 (96.9\%)\\
        \hline
       \end{tabular}
       \caption{(A) Lorenz 96 systems, all coordinates}
    \end{subtable}
    \vfill
    \vfill
    \begin{subtable}[h]{\textwidth}
        \centering
        \begin{tabular}{l r r}
         & Coupling (detected) & No coupling (detected) \\
        \hline
        Coupling (real) & 25/30 (83.3\%) & 5/30 (16.7\%)\\
        No coupling (real) & 0/32 (0\%) & 32/32 (100\%)\\
        \hline
        \end{tabular}
        \caption{(B) Lorenz 96 systems, time-delay embeddings}
     \end{subtable}
     \caption{Confusion matrices for the outcomes of the significance test applied to the unidirectionally coupled Lorenz 96 systems, (\emph{A}) using all coordinates and (\emph{B}) using time-delay embeddings.
     The lines of the tables report the true causal scenario (presence or absence of the coupling) while the columns refer to the outcomes of the hypothesis testing. 
     Null distributions with 1000 samples were employed and the null hypothesis (absence of coupling) was rejected for $p < 0.001$.
     The class ``no coupling'' contains two more elements than the class ``coupling'' because it includes, for the particular coupling $\varepsilon = 0$, both the directions $X\rightarrow Y$ and $Y\rightarrow X$.}
     \label{tab:confusion_matrix}
\end{table}

\clearpage
\printbibliography[keyword={SI}, resetnumbers=true, title={Supplementary references}]

@article{glielmo2022dadapy,
title = {DADApy: Distance-based analysis of data-manifolds in Python},
journal = {Patterns},
volume = {3},
keywords = {main-text},
%number = {10},
eid = {100589},
%pages = {100589},
year = {2022},
issn = {2666-3899},
doi = {https://doi.org/10.1016/j.patter.2022.100589},
url = {https://www.sciencedirect.com/science/article/pii/S2666389922002070},
author = {Aldo Glielmo and Iuri Macocco and Diego Doimo and Matteo Carli and Claudio Zeni and Romina Wild and Maria d’Errico and Alex Rodriguez and Alessandro Laio},
}

@article{glielmo2022ranking,
    author = {Glielmo, Aldo and Zeni, Claudio and Cheng, Bingqing and Csányi, Gábor and Laio, Alessandro},
    title = "{Ranking the information content of distance measures}",
    journal = {PNAS Nexus},
    volume = {1},
    keywords = {main-text},
    eid = {2752-6542},
    %number = {2},
    year = {2022},
    issn = {2752-6542},
    doi = {10.1093/pnasnexus/pgac039},
    url = {https://doi.org/10.1093/pnasnexus/pgac039},
    %note = {pgac039},
    eprint = {https://academic.oup.com/pnasnexus/article-pdf/1/2/pgac039/44246399/pgac039.pdf},
}

@book{pearl2009causality,
author = {Pearl, Judea},
title = {Causality: Models, Reasoning and Inference},
year = {2009},
isbn = {052189560X},
publisher = {Cambridge University Press},
keywords = {main-text},
%address = {USA},
%edition = {2nd},
}

@article{chicharro2009reliable,
  title = {Reliable detection of directional couplings using rank statistics},
  author = {Chicharro, Daniel and Andrzejak, Ralph G.},
  journal = {Phys. Rev. E},
  volume = {80},
  keywords = {main-text},
  %issue = {2},
  eid = {026217},
  %pages = {026217},
  %numpages = {5},
  year = {2009},
  publisher = {American Physical Society},
  doi = {10.1103/PhysRevE.80.026217},
  url = {https://link.aps.org/doi/10.1103/PhysRevE.80.026217},
}

@article{palus2007directionality,
  title = {Directionality of coupling from bivariate time series: How to avoid false causalities and missed connections},
  author = {Palu\ifmmode\check{s}\else\v{s}\fi{}, Milan and Vejmelka, Martin},
  journal = {Phys. Rev. E},
  volume = {75},
  keywords = {main-text},
  %issue = {5},
  eid = {056211},
  %pages = {056211},
  numpages = {14},
  year = {2007},
  publisher = {American Physical Society},
  doi = {10.1103/PhysRevE.75.056211},
  url = {https://link.aps.org/doi/10.1103/PhysRevE.75.056211},
}

@article{palus2018causality,
author = {Paluš,Milan  and Krakovská,Anna  and Jakubík,Jozef  and Chvosteková,Martina },
title = {Causality, dynamical systems and the arrow of time},
journal = {Chaos},%: An Interdisciplinary Journal of Nonlinear Science}

@article{boccaletti2002thesync,
  title={The synchronization of chaotic systems},
  author={Stefano Boccaletti and J{\"u}rgen Kurths and Grigory V. Osipov and Diego L. Valladares and Changsong Zhou},
  journal={Physics Reports},
  year={2002},
  volume={366},
  eid = {1-101},
  %pages={1-101},
  doi = {10.1016/S0370-1573(02)00137-0},
keywords = {main-text},
}

@book{imbens_rubin_2015, place={Cambridge}, title={Causal Inference for Statistics, Social, and Biomedical Sciences: An Introduction}, DOI={10.1017/CBO9781139025751}, publisher={Cambridge University Press}, author={Imbens, Guido W. and Rubin, Donald B.}, year={2015},
}

@article{krakovska2018comparison,
  title = {Comparison of six methods for the detection of causality in a bivariate time series},
  author = {Krakovsk\'a, Anna and Jakub\'{\i}k, Jozef and Chvostekov\'a, Martina and Coufal, David and Jajcay, Nikola and Palu\ifmmode \check{s}\else \v{s}\fi{}, Milan},
  journal = {Phys. Rev. E},
  volume = {97},
  eid = {042207},
  %issue = {4},
  %pages = {042207},
  numpages = {14},
  year = {2018},
  %month = {Apr},
  publisher = {American Physical Society},
  doi = {10.1103/PhysRevE.97.042207},
  url = {https://link.aps.org/doi/10.1103/PhysRevE.97.042207},
  keywords = {main-text},
}

@article{leng2020partialCM,
  title={Partial cross mapping eliminates indirect causal influences},
  author={Siyang Leng and Huanfei Ma and J{\"u}rgen Kurths and Ying-Cheng Lai and Wei Lin and Kazuyuki Aihara and Luonan Chen},
  journal={Nature Communications},
  year={2020},
  volume={11},
  url={https://api.semanticscholar.org/CorpusID:256637650},
}

@article {lorenz1963deterministic,
      author = "Edward N.  Lorenz",
      title = "Deterministic Nonperiodic Flow",
      journal = "Journal of Atmospheric Sciences",
      year = "1963",
      publisher = "American Meteorological Society",
      address = "Boston MA, USA",
      volume = "20",
      %number = "2",
      doi = "10.1175/1520-0469(1963)020<0130:DNF>2.0.CO;2",
      eid = "130-141",
      %pages=      "130 - 141",
      url = "https://journals.ametsoc.org/view/journals/atsc/20/2/1520-0469_1963_020_0130_dnf_2_0_co_2.xml",
keywords = {main-text},
}

@inbook{lorenz2006predictability, place={Cambridge}, title={Predictability – a problem partly solved}, DOI={10.1017/CBO9780511617652.004}, booktitle={Predictability of Weather and Climate}, publisher={Cambridge University Press}, author={Lorenz, Edward N.}, editor={Palmer, Tim and Hagedorn, RenateEditors}, year={2006}, pages={40–58},
}

@article{mooij2016distinguishing,
author = {Mooij, Joris M. and Peters, Jonas and Janzing, Dominik and Zscheischler, Jakob and Sch\"{o}lkopf, Bernhard},
title = {Distinguishing Cause from Effect Using Observational Data: Methods and Benchmarks},
year = {2016},
issue_date = {January 2016},
publisher = {JMLR.org},
volume = {17},
%number = {1},
eid = {1103–1204},
issn = {1532-4435},
journal = {J. Mach. Learn. Res.},
%pages = {1103–1204},
url = {http://jmlr.org/papers/v17/14-518.html},
keywords = {main-text},
}

@article{springer2023data,
title = {Data based quantification of synchronization},
journal = {Foundations of Data Science},
volume = {5},
%number = {1},
%pages = {152-176},
eid = {152-176},
year = {2023},
issn = {},
doi = {10.3934/fods.2022020},
url = {https://www.aimsciences.org/article/id/6375fe4e316e21069b91345e},
author = {Dipal Shah and Sebastian Springer and Heikki Haario and Bernardo Barbiellini and Leonid Kalachev},
keywords = {Synchronization, quasi-complete synchronization, Chaos suppression, correlation integral likelihood, feature vector, emerging synchronization, parameter estimation},
keywords = {main-text},
}

@article{rossler1976anequation,
title = {An equation for continuous chaos},
journal = {Physics Letters A},
volume = {57},
%number = {5},
eid = {397-398},
%pages = {397-398},
year = {1976},
issn = {0375-9601},
doi = {10.1016/0375-9601(76)90101-8},
url = {https://www.sciencedirect.com/science/article/pii/0375960176901018},
author = {O.E. Rössler},
abstract = {A prototype equation to the Lorenz model of turbulence contains just one (second-order) nonlinearity in one variable. The flow in state space allows for a “folded” Poincaré map (horseshoe map). Many more natural and artificial systems are governed by this type of equation.},
keywords = {main-text},
}

@article{krakovska2016testing,
  title = {Testing for causality in reconstructed state spaces by an optimized mixed prediction method},
  author = {Krakovsk\'a, Anna and Hanzely, Filip},
  journal = {Phys. Rev. E},
  volume = {94},
  %issue = {5},
  eid = {052203},
  %pages = {052203},
  numpages = {9},
  year = {2016},
  publisher = {American Physical Society},
  doi = {10.1103/PhysRevE.94.052203},
  url = {https://link.aps.org/doi/10.1103/PhysRevE.94.052203},
  keywords = {main-text},
}

@article{runge2018causal,
    author = {Runge, J.},
    title = "{Causal network reconstruction from time series: From theoretical assumptions to practical estimation}",
    journal = {Chaos},%: An Interdisciplinary Journal of Nonlinear Science}

@article{krakovska2022state,
title = {State space reconstruction techniques and the accuracy of prediction},
journal = {Commun. Nonlinear Sci. Numer. Simul.},%Communications in Nonlinear Science and Numerical Simulation}

@article{schreiber2000measuring,
  title = {Measuring Information Transfer},
  author = {Schreiber, Thomas},
  journal = {Phys. Rev. Lett.},
  volume = {85},
  %issue = {2},
  eid = {461--464},
  %pages = {461--464},
  numpages = {0},
  year = {2000},
  publisher = {American Physical Society},
  doi = {10.1103/PhysRevLett.85.461},
  url = {https://link.aps.org/doi/10.1103/PhysRevLett.85.461},
  keywords = {main-text},
}

@article{palus2001synchronization,
  title = {Synchronization as adjustment of information rates: Detection from bivariate time series},
  author = {Palu\ifmmode\check{s}\else\v{s}\fi{}, Milan and Kom\'arek, Vladim\'{\i}r and Hrn\ifmmode\check{c}\else \v{c}\fi{}\'{\i}\ifmmode \check{r}\else\v{r}\fi{}, Zbyn\ifmmode \check{e}\else\v{e}\fi{}k and \ifmmode\check{S}\else \v{S}\fi{}t\ifmmode \check{e}\else\v{e}\fi{}rbov\'a, Katalin},
  journal = {Phys. Rev. E},
  volume = {63},
  %issue = {4},
  eid = {046211},
  %pages = {046211},
  numpages = {6},
  year = {2001},
  publisher = {American Physical Society},
  doi = {10.1103/PhysRevE.63.046211},
  url = {https://link.aps.org/doi/10.1103/PhysRevE.63.046211},
  keywords = {main-text},
}

@article{barnett2009granger,
  title = {Granger Causality and Transfer Entropy Are Equivalent for Gaussian Variables},
  author = {Barnett, Lionel and Barrett, Adam B. and Seth, Anil K.},
  journal = {Phys. Rev. Lett.},
  volume = {103},
  %issue = {23},
  eid = {238701},
  %pages = {238701},
  numpages = {4},
  year = {2009},
  publisher = {American Physical Society},
  doi = {10.1103/PhysRevLett.103.238701},
  url = {https://link.aps.org/doi/10.1103/PhysRevLett.103.238701},
  keywords = {main-text},
}

@article{granger1969investigating,
 ISSN = {00129682, 14680262},
 URL = {http://www.jstor.org/stable/1912791},
 author = {C. W. J. Granger},
 journal = {Econometrica},
 %number = {3},
 eid = {424--438},
 %pages = {424--438},
 publisher = {[Wiley, Econometric Society]},
 title = {Investigating Causal Relations by Econometric Models and Cross-spectral Methods},
 urldate = {2022-11-07},
 volume = {37},
 year = {1969},
 keywords = {main-text},
}

@BOOK{beckenbach1956modern,
  TITLE = {N. Wiener, in: Modern Mathematics for Engineers},
  SUBTITLE = {The Science of Microfabrication},
  AUTHOR = {Beckenbach, Edwin},
  YEAR = {1956}, 
  PUBLISHER = {McGraw-Hill, New York},
  %CHAPTER = {8},
  keywords = {main-text},
}

@article{Chen2004analyzing,
title = {Analyzing multiple nonlinear time series with extended Granger causality},
journal = {Phys. Lett. A},%Physics Letters A}

@article{
sugihara2012detecting,
author = {George Sugihara  and Robert May  and Hao Ye  and Chih-hao Hsieh  and Ethan Deyle  and Michael Fogarty  and Stephan Munch },
title = {Detecting Causality in Complex Ecosystems},
journal = {Science},
volume = {338},
%number = {6106},
eid = {496-500},
%pages = {496-500},
year = {2012},
doi = {10.1126/science.1227079},
URL = {https://www.science.org/doi/abs/10.1126/science.1227079},
eprint = {https://www.science.org/doi/pdf/10.1126/science.1227079},
keywords = {main-text},
}

@InProceedings{takens1981dynamical,
author="Takens, Floris",
editor="Rand, David and Young, Lai-Sang",
title="Detecting strange attractors in turbulence",
booktitle="Dynamical Systems and Turbulence, Warwick 1980",
year="1981",
publisher="Springer Berlin Heidelberg",
%address="Berlin, Heidelberg",
%eid="366--381",
%pages="366--381",
isbn="978-3-540-38945-3",
keywords = {main-text},
}

@article{ofir2022neural,
title = {Neural signatures of evidence accumulation in temporal decisions},
journal = {Curr. Biol.},%Current Biology}

@article{kennel1992determining,
  title = {Determining embedding dimension for phase-space reconstruction using a geometrical construction},
  author = {Kennel, Matthew B. and Brown, Reggie and Abarbanel, Henry D. I.},
  journal = {Phys. Rev. A},
  volume = {45},
  %issue = {6},
  eid = {3403--3411},
  %pages = {3403--3411},
  numpages = {0},
  year = {1992},
  publisher = {American Physical Society},
  doi = {10.1103/PhysRevA.45.3403},
  url = {https://link.aps.org/doi/10.1103/PhysRevA.45.3403},
  keywords = {main-text},
}

@article{marinazzo2008kernel,
  title = {Kernel Method for Nonlinear Granger Causality},
  author = {Marinazzo, Daniele and Pellicoro, Mario and Stramaglia, Sebastiano},
  journal = {Phys. Rev. Lett.},
  volume = {100},
  %issue = {14},
  eid = {144103},  
  %pages = {144103},
  numpages = {4},
  year = {2008},
  publisher = {American Physical Society},
  doi = {10.1103/PhysRevLett.100.144103},
  url = {https://link.aps.org/doi/10.1103/PhysRevLett.100.144103},
  keywords = {main-text},
}

@article{marinazzo2006nonlinear,
  title = {Nonlinear parametric model for Granger causality of time series},
  author = {Marinazzo, Daniele and Pellicoro, Mario and Stramaglia, Sebastiano},
  journal = {Phys. Rev. E},
  volume = {73},
  %issue = {6},
  %pages = {066216},
  eid = {066216},
  numpages = {6},
  year = {2006},
  publisher = {American Physical Society},
  doi = {10.1103/PhysRevE.73.066216},
  url = {https://link.aps.org/doi/10.1103/PhysRevE.73.066216},
  keywords = {main-text},
}

@article{barnett2015granger,
  title = {Granger causality for state-space models},
  author = {Barnett, Lionel and Seth, Anil K.},
  journal = {Phys. Rev. E},
  volume = {91},
  %issue = {4},
  eid = {040101},
  %pages = {040101},
  numpages = {5},
  year = {2015},
  publisher = {American Physical Society},
  doi = {10.1103/PhysRevE.91.040101},
  url = {https://link.aps.org/doi/10.1103/PhysRevE.91.040101},
  keywords = {main-text},
}

@Inbook{hoover2017causality,
author="Hoover, Kevin D.",
title="Causality in Economics and Econometrics",
bookTitle="The New Palgrave Dictionary of Economics",
year="2017",
publisher="Palgrave Macmillan UK",
%address="London",
%eid="1--13",
%pages="1--13",
isbn="978-1-349-95121-5",
doi="10.1057/978-1-349-95121-5_2227-1",
url="https://doi.org/10.1057/978-1-349-95121-5_2227-1",
keywords = {main-text},
}

@article {kretschmer2016using,
  author = "Marlene Kretschmer and Dim Coumou and Jonathan F. Donges and Jakob Runge",
  title = "Using Causal Effect Networks to Analyze Different Arctic Drivers of Midlatitude Winter Circulation",
  journal = "J. Clim.",%Journal of Climate",
  year = "2016",
  publisher = "American Meteorological Society",
  address = "Boston MA, USA",
  volume = "29",
  %number = "11",
  doi = "https://doi.org/10.1175/JCLI-D-15-0654.1",
  eid=      "4069-4081",
  %pages=   "4069-4081",
  url = "https://journals.ametsoc.org/view/journals/clim/29/11/jcli-d-15-0654.1.xml",
  keywords = {main-text},
}

@article{friston2003dynamic,
title = {Dynamic causal modelling},
journal = {NeuroImage},
volume = {19},
%number = {4},
eid = {1273-1302},
%pages = {1273-1302},
year = {2003},
issn = {1053-8119},
doi = {https://doi.org/10.1016/S1053-8119(03)00202-7},
url = {https://www.sciencedirect.com/science/article/pii/S1053811903002027},
author = {K.J. Friston and L. Harrison and W. Penny},
keywords = {main-text},
}

@article{kaminski2001evaluating,
  title={Evaluating causal relations in neural systems: Granger causality, directed transfer function and statistical assessment of significance},
  author={Maciej Kaminski and Mingzhou Ding and Wilson A. Truccolo and Steven L. Bressler},
  journal={Biol. Cybern.},%Biological Cybernetics}

@article{damsma2021temporal,
	author = {Damsma, Atser and Schlichting, Nadine and van Rijn, Hedderik},
	title = {Temporal Context Actively Shapes EEG Signatures of Time Perception},
	volume = {41},
	%number = {20},
	eid = {4514--4523},
	%pages = {4514--4523},
	year = {2021},
	doi = {10.1523/JNEUROSCI.0628-20.2021},
	publisher = {Society for Neuroscience},
	issn = {0270-6474},
	URL = {https://www.jneurosci.org/content/41/20/4514},
	eprint = {https://www.jneurosci.org/content/41/20/4514.full.pdf},
	journal = {J. Neurosci.},%Journal of Neuroscience}

@misc{gendron2023survey,
      title={A Survey of Methods, Challenges and Perspectives in Causality}, 
      author={Gaël Gendron and Michael Witbrock and Gillian Dobbie},
      year={2023},
      eprint={2302.00293},
      archivePrefix={arXiv},
      primaryClass={cs.LG},
      journal = {ArXiv},
      eid = {2302.00293},
}

@article{spirtes2016causal,
  title = {Causal discovery and inference: concepts and recent methodological advances},
  author = {Spirtes, P. and Zhang, K.},
  journal = {Appl. Inform.},%Applied Informatics}

@ARTICLE{runge2019inferring,
       author = {{Runge}, Jakob and {Bathiany}, Sebastian and {Bollt}, Erik and {Camps-Valls}, Gustau and {Coumou}, Dim and {Deyle}, Ethan and {Glymour}, Clark and {Kretschmer}, Marlene and {Mahecha}, Miguel D. and {Mu{\~n}oz-Mar{\'\i}}, Jordi and {van Nes}, Egbert H. and {Peters}, Jonas and {Quax}, Rick and {Reichstein}, Markus and {Scheffer}, Marten and {Sch{\"o}lkopf}, Bernhard and {Spirtes}, Peter and {Sugihara}, George and {Sun}, Jie and {Zhang}, Kun and {Zscheischler}, Jakob},
        title = "{Inferring causation from time series in Earth system sciences}",
      journal = {Nat. Comm.},%Nature Communications}

@article {yuan2022datadriven,
article_type = {journal},
title = {Data-driven causal analysis of observational biological time series},
author = {Yuan, Alex Eric and Shou, Wenying},
%editor = {Schuman, Meredith C},
volume = 11,
year = 2022,
pub_date = {2022-08-19},
eid = {e72518},
%pages = {e72518},
citation = {eLife 2022;11:e72518},
doi = {10.7554/eLife.72518},
url = {https://doi.org/10.7554/eLife.72518},
keywords = {time series, causality, model-free, surrogate data, convergent cross-mapping, Granger causality},
journal = {eLife},
issn = {2050-084X},
publisher = {eLife Sciences Publications, Ltd},
keywords = {main-text},
}

@article{kononowicz2014decoupling,
  title={Decoupling Interval Timing and Climbing Neural Activity: A Dissociation between CNV and N1P2 Amplitudes},
  author={Tadeusz Władysław Kononowicz and Hedderik van Rijn},
  journal={J. Neurosci.},%The Journal of Neuroscience}

@article{tonoyan2022time,
title = {Subjective time is predicted by local and early visual processing},
journal = {NeuroImage},
volume = {264},
eid = {119707},
%pages = {119707},
year = {2022},
issn = {1053-8119},
doi = {https://doi.org/10.1016/j.neuroimage.2022.119707},
url = {https://www.sciencedirect.com/science/article/pii/S105381192200828X},
author = {Yelena Tonoyan and Michele Fornaciai and Brent Parsons and Domenica Bueti},
keywords = {Time perception, Motion adaptation, Duration compression, EEG, Neural decoding},
keywords = {main-text},
}

@article{runge2012escaping,
  title = {Escaping the Curse of Dimensionality in Estimating Multivariate Transfer Entropy},
  author = {Runge, Jakob and Heitzig, Jobst and Petoukhov, Vladimir and Kurths, J\"urgen},
  journal = {Phys. Rev. Lett.},
  volume = {108},
  %issue = {25},
  eid = {258701},
  %pages = {258701},
  numpages = {5},
  year = {2012},
  publisher = {American Physical Society},
  doi = {10.1103/PhysRevLett.108.258701},
  url = {https://link.aps.org/doi/10.1103/PhysRevLett.108.258701},
  keywords = {main-text},
}

@article{bressler2011wiener,
title = {Wiener–Granger Causality: A well established methodology},
journal = {NeuroImage},
volume = {58},
%number = {2},
eid = {323-329},
%pages = {323-329},
year = {2011},
issn = {1053-8119},
doi = {https://doi.org/10.1016/j.neuroimage.2010.02.059},
url = {https://www.sciencedirect.com/science/article/pii/S1053811910002272},
author = {Steven L. Bressler and Anil K. Seth},
keywords = {Autoregressive model, Brain, Causality, Inference, Neuroscience, Time series},
keywords = {main-text},
}

@article{wang2022newmethod,
title = {A new method of nonlinear causality detection: Reservoir computing Granger causality},
journal = {Chaos Solitons \& Fractals},%Chaos, Solitons \& Fractals}

@article{Thorpe1996,
author = {Thorpe, Simon and Fize, Denis and Marlot, Catherine},
year = {1996},
eid = {520-2},
%pages = {520-2},
title = {Speed of Processing in the Human Visual System},
volume = {381},
journal = {Nature},
doi = {10.1038/381520a0},
keywords = {main-text},
}

@article{Fabre-Thorpe2001,
    author = {Fabre-Thorpe, Michèle and Delorme, Arnaud and Marlot, Catherine and Thorpe, Simon},
    title = "{A Limit to the Speed of Processing in Ultra-Rapid Visual Categorization of Novel Natural Scenes}",
    journal = {J. Cogn. Neurosci.},%Journal of Cognitive Neuroscience}

@inproceedings{seabold2010statsmodels,
  title={Statsmodels: Econometric and statistical modeling with python},
  author={Seabold, Skipper and Perktold, Josef},
  booktitle={9th Python in Science Conference},
  year={2010},
}

@ARTICLE{virtanen2020SciPy,
  author  = {Virtanen, Pauli and Gommers, Ralf and Oliphant, Travis E. and
            Haberland, Matt and Reddy, Tyler and Cournapeau, David and
            Burovski, Evgeni and Peterson, Pearu and Weckesser, Warren and
            Bright, Jonathan and {van der Walt}, St{\'e}fan J. and
            Brett, Matthew and Wilson, Joshua and Millman, K. Jarrod and
            Mayorov, Nikolay and Nelson, Andrew R. J. and Jones, Eric and
            Kern, Robert and Larson, Eric and Carey, C J and
            Polat, {\.I}lhan and Feng, Yu and Moore, Eric W. and
            {VanderPlas}, Jake and Laxalde, Denis and Perktold, Josef and
            Cimrman, Robert and Henriksen, Ian and Quintero, E. A. and
            Harris, Charles R. and Archibald, Anne M. and
            Ribeiro, Ant{\^o}nio H. and Pedregosa, Fabian and
            {van Mulbregt}, Paul and {SciPy 1.0 Contributors}},
  title   = {{{SciPy} 1.0: Fundamental Algorithms for Scientific
            Computing in Python}},
  journal = {Nat. Methods},%Nature Methods}

@article{Protopapa2019,
    doi = {10.1371/journal.pbio.3000026},
    author = {Protopapa, Foteini AND Hayashi, Masamichi J. AND Kulashekhar, Shrikanth AND van der Zwaag, Wietske AND Battistella, Giovanni AND Murray, Micah M. AND Kanai, Ryota AND Bueti, Domenica},
    journal = {PLoS Biol.},%PLOS Biology}

@article{Protopapa2023,
author = {Protopapa, Foteini and Kulashekhar, Shrikanth and Hayashi, Masamichi J. and Kanai, Ryota and Bueti, Domenica},
title = {{Effective connectivity in a duration selective cortico-cerebellar network}},
journal = {Sci. Rep.},
volume = {13},
number = {20674},
pages = {1--17},
year = {2023},
month = nov,
issn = {2045-2322},
publisher = {Nature Publishing Group},
doi = {10.1038/s41598-023-47954-4},
}

@article{castro2023timeseries,
  title={Time series causal relationships discovery through feature importance and ensemble models},
  author={Castro, M. and Mendes Júnior, P.R. and Soriano-Vargas, A. and others},
  journal={Scientific Reports},
  volume={13},
  pages={11402},
  year={2023},
  publisher={Nature Publishing Group},
  doi={10.1038/s41598-023-37929-w},
}

@article{dirusso2002cortical_SI,
author = {Di Russo, Francesco and Martínez, Antígona and Sereno, Martin I. and Pitzalis, Sabrina and Hillyard, Steven A.},
title = {Cortical sources of the early components of the visual evoked potential},
journal = {Human Brain Mapping},
volume = {15},
number = {2},
pages = {95-111},
doi = {https://doi.org/10.1002/hbm.10010},
url = {https://onlinelibrary.wiley.com/doi/abs/10.1002/hbm.10010},
eprint = {https://onlinelibrary.wiley.com/doi/pdf/10.1002/hbm.10010},
year = {2002}
}

@book{casella2001statistical_SI,
  author = {Casella, George and Berger, Roger},
  publisher = {{Duxbury Resource Center}},
  title = {Statistical Inference},
  year = {2001},
}

@BOOK{nelsen2006anintroduction_SI,
  TITLE = {An introduction to copulas},
  SUBTITLE = {The Science of Microfabrication},
  AUTHOR = {Nelsen, Roger B},
  YEAR = {2006}, 
  PUBLISHER = {Springer, New York},
  doi = {10.1007/0-387-28678-0},
}

@article{calsaverini2009aninformation_SI,
	doi = {10.1209/0295-5075/88/68003},
	url = {https://doi.org/10.1209/0295-5075/88/68003},
	year = 2009,
	publisher = {{IOP} Publishing},
	volume = {88},
	%number = {6},
	eid = {68003},
	%pages = {68003},
	author = {R. S. Calsaverini and R. Vicente},
	title = {An information-theoretic approach to statistical dependence: Copula information},
	journal = {Europhys. Lett.},%{EPL} (Europhysics Letters)}

@article{safaai2018information_SI,
  title = {Information estimation using nonparametric copulas},
  author = {Safaai, Houman and Onken, Arno and Harvey, Christopher D. and Panzeri, Stefano},
  journal = {Phys. Rev. E},
  volume = {98},
  %issue = {5},
  eid = {68003},
  %pages = {053302},
  %numpages = {15},
  year = {2018},
  publisher = {American Physical Society},
  doi = {10.1103/PhysRevE.98.053302},
  url = {https://link.aps.org/doi/10.1103/PhysRevE.98.053302},
  keywords = {SI},
}

@ARTICLE{dowson1973maxent_SI,
  author={Dowson, D. and Wragg, A.},
  journal={IEEE Transactions on Information Theory}, 
  title={Maximum-entropy distributions having prescribed first and second moments (Corresp.)}, 
  year={1973},
  volume={19},
  %number={5},
  %pages={689-693},
  eid={689-693},
  doi={10.1109/TIT.1973.1055060},
keywords = {SI},
}

@article{DELORME20049_SI,
title = {EEGLAB: an open source toolbox for analysis of single-trial EEG dynamics including independent component analysis},
journal = {J. Neurosci. Methods},%Journal of Neuroscience Methods}

@article{PIONTONACHINI2019181_SI,
title = {ICLabel: An automated electroencephalographic independent component classifier, dataset, and website},
journal = {NeuroImage},
volume = {198},
eid = {181-197},
%pages = {181-197},
year = {2019},
issn = {1053-8119},
doi = {https://doi.org/10.1016/j.neuroimage.2019.05.026},
url = {https://www.sciencedirect.com/science/article/pii/S1053811919304185},
author = {Luca Pion-Tonachini and Ken Kreutz-Delgado and Scott Makeig},
keywords = {SI},
}

@article{
sugihara2012detecting_SI,
author = {George Sugihara  and Robert May  and Hao Ye  and Chih-hao Hsieh  and Ethan Deyle  and Michael Fogarty  and Stephan Munch },
title = {Detecting Causality in Complex Ecosystems},
journal = {Science},
volume = {338},
%number = {6106},
eid = {496-500},
%pages = {496-500},
year = {2012},
doi = {10.1126/science.1227079},
URL = {https://www.science.org/doi/abs/10.1126/science.1227079},
eprint = {https://www.science.org/doi/pdf/10.1126/science.1227079},
keywords = {SI},
}

@article{chicharro2009reliable_SI,
  title = {Reliable detection of directional couplings using rank statistics},
  author = {Chicharro, Daniel and Andrzejak, Ralph G.},
  journal = {Phys. Rev. E},
  volume = {80},
  keywords = {SI},
  %issue = {2},
  eid = {026217},
  %pages = {026217},
  %numpages = {5},
  year = {2009},
  publisher = {American Physical Society},
  doi = {10.1103/PhysRevE.80.026217},
  url = {https://link.aps.org/doi/10.1103/PhysRevE.80.026217},
}

@article{Chen2004analyzing_SI,
title = {Analyzing multiple nonlinear time series with extended Granger causality},
journal = {Phys. Lett. A},%Physics Letters A}

@article{fraser1986independent_SI,
  title = {Independent coordinates for strange attractors from mutual information},
  author = {Fraser, Andrew M. and Swinney, Harry L.},
  journal = {Phys. Rev. A},
  volume = {33},
  %issue = {2},
  eid = {1134--1140},
  %pages = {1134--1140},
  %numpages = {0},
  year = {1986},
  publisher = {American Physical Society},
  doi = {10.1103/PhysRevA.33.1134},
  url = {https://link.aps.org/doi/10.1103/PhysRevA.33.1134},
  keywords = {SI},
}

@article{glielmo2022ranking_SI,
    author = {Glielmo, Aldo and Zeni, Claudio and Cheng, Bingqing and Csányi, Gábor and Laio, Alessandro},
    title = "{Ranking the information content of distance measures}",
    journal = {PNAS Nexus},
    volume = {1},
    keywords = {SI},
    eid = {2752-6542},
    %number = {2},
    year = {2022},
    issn = {2752-6542},
    doi = {10.1093/pnasnexus/pgac039},
    url = {https://doi.org/10.1093/pnasnexus/pgac039},
    %note = {pgac039},
    eprint = {https://academic.oup.com/pnasnexus/article-pdf/1/2/pgac039/44246399/pgac039.pdf},
}

@article{krakovska2018comparison_SI,
  title = {Comparison of six methods for the detection of causality in a bivariate time series},
  author = {Krakovsk\'a, Anna and Jakub\'{\i}k, Jozef and Chvostekov\'a, Martina and Coufal, David and Jajcay, Nikola and Palu\ifmmode \check{s}\else \v{s}\fi{}, Milan},
  journal = {Phys. Rev. E},
  volume = {97},
  eid = {042207},
  %issue = {4},
  %pages = {042207},
  numpages = {14},
  year = {2018},
  %month = {Apr},
  publisher = {American Physical Society},
  doi = {10.1103/PhysRevE.97.042207},
  url = {https://link.aps.org/doi/10.1103/PhysRevE.97.042207},
  keywords = {SI},
}

@article{mesner2021conditional_SI,
  author={Mesner, Octavio César and Shalizi, Cosma Rohilla},
  journal={IEEE Transactions on Information Theory}, 
  title={Conditional Mutual Information Estimation for Mixed, Discrete and Continuous Data}, 
  year={2021},
  volume={67},
  number={1},
  pages={464-484},
  doi={10.1109/TIT.2020.3024886},
}

@article{palus2007directionality_SI,
  title = {Directionality of coupling from bivariate time series: How to avoid false causalities and missed connections},
  author = {Palu\ifmmode\check{s}\else\v{s}\fi{}, Milan and Vejmelka, Martin},
  journal = {Phys. Rev. E},
  volume = {75},
  keywords = {SI},
  %issue = {5},
  eid = {056211},
  %pages = {056211},
  numpages = {14},
  year = {2007},
  publisher = {American Physical Society},
  doi = {10.1103/PhysRevE.75.056211},
  url = {https://link.aps.org/doi/10.1103/PhysRevE.75.056211},
}

@article{runge2018causal_SI,
    author = {Runge, J.},
    title = "{Causal network reconstruction from time series: From theoretical assumptions to practical estimation}",
    journal = {Chaos},%: An Interdisciplinary Journal of Nonlinear Science}

\end{document}